\documentclass[aps,prb,reprint,twocolumn,superscriptaddress,amsmath,amssymb]{revtex4}
\usepackage{graphicx}%
\usepackage{float}
\usepackage[T1]{fontenc}
\usepackage{xcolor}
\usepackage{ulem}
\usepackage[colorlinks=true, urlcolor=blue, linkcolor=blue, citecolor=blue, bookmarks, pdftitle={article}]{hyperref}
\usepackage[]{units}
\usepackage{times}

\usepackage{bm}

\newcommand{\comment}[1]{}

\newcommand{\e}{\mathrm{e}}
\renewcommand{\d}{\mathrm d}
\let\ifr\i
\renewcommand{\i}{\mathrm{i}}

\newcommand{\blue}[1]{{#1}} 
\newcommand{\red}[1]{} 

\renewcommand{\emph}{\textit}

\let\oldaddcontentsline\addcontentsline
\newcommand{\stoptocentries}{\renewcommand{\addcontentsline}[3]{}}
\newcommand{\starttocentries}{\let\addcontentsline\oldaddcontentsline}

\begin{document}
\title{Valley-magnetophonon resonance for interlayer excitons}
\author{Dmitry S. Smirnov}
\email{smirnov@mail.ioffe.ru}
\affiliation{Ioffe Institute, 194021 St. Petersburg, Russia}
\author{Johannes Holler}
\affiliation{Institut f\"ur Experimentelle und Angewandte Physik,
	Universit\"at Regensburg, 93040 Regensburg, Germany}

\author{Michael Kempf}
\affiliation{Institut f\"ur  Physik, Universit\"at Rostock, 18059 Rostock, Germany}	

\author{Jonas Zipfel}
\affiliation{Institut f\"ur Experimentelle und Angewandte Physik, Universit\"at Regensburg, 93040 Regensburg, Germany}
\affiliation{Molecular Foundry, Lawrence Berkeley National Laboratory, Berkeley, California 94720, USA}

\author{Philipp Nagler}
\affiliation{Institut f\"ur Experimentelle und Angewandte Physik, Universit\"at Regensburg, 93040 Regensburg, Germany}
	
\author{Mariana V. Ballottin}
\affiliation{High Field Magnet Laboratory (HFML - EMFL), Radboud University, 6525 ED Nijmegen, The Netherlands}

\author{Anatolie A. Mitioglu}
\affiliation{High Field Magnet Laboratory (HFML - EMFL), Radboud University, 6525 ED Nijmegen, The Netherlands}

\author{Alexey Chernikov}
\affiliation{Institut f\"ur Experimentelle und Angewandte Physik,
	Universit\"at Regensburg, 93040 Regensburg, Germany}
\affiliation{Dresden Integrated Center for Applied Physics and Photonic Materials (IAPP) and W\"urzburg-Dresden Cluster of Excellence ct.qmat, Technische Universit\"at Dresden, 01062 Dresden, Germany}

\author{Peter C. M. Christianen}
\affiliation{High Field Magnet Laboratory (HFML - EMFL), Radboud University, 6525 ED Nijmegen, The Netherlands}

\author{Christian Sch\"uller}
\affiliation{Institut f\"ur Experimentelle und Angewandte Physik,
	Universit\"at Regensburg, 93040 Regensburg, Germany}

\author{Tobias Korn}
\affiliation{Institut f\"ur  Physik, Universit\"at Rostock, 18059 Rostock, Germany}

\begin{abstract}
  Heterobilayers consisting of MoSe$_2$ and WSe$_2$ monolayers can host optically bright interlayer excitons with intriguing properties such as ultralong lifetimes and pronounced circular polarization of their photoluminescence due to valley polarization, which can be induced by circularly polarized excitation or applied magnetic fields. Here, we report on the observation of an intrinsic  valley-magnetophonon resonance for localized interlayer excitons promoted by invervalley hole scattering. It leads to a resonant increase of the photoluminescence polarization degree at the same field of 24.2~Tesla for H-type and R-type stacking configurations despite their vastly different excitonic energy splittings. As a microscopic mechanism of the hole intervalley scattering we identify the scattering with chiral TA phonons of MoSe$_2$ between excitonic states mixed by the long-range electron hole exchange interaction.
\end{abstract}

\maketitle

\stoptocentries
\section{Introduction}
\starttocentries

Two-dimensional (2D) crystals and their van der Waals (vdW) heterostructures (HS) are promising candidates for novel optoelectronic devices. Among the 2D crystals, the semiconducting transition metal dichalcogenides (TMDCs) like MoS$_2$ have garnered a lot of attention due their intriguing properties: in the monolayer (ML) limit, they are direct-gap semiconductors~\cite{Mak2010,Splendiani2010} with large exciton binding energies~\cite{Chernikov2014} and peculiarities such as spin-valley locking~\cite{Xiao2012}. The latter phenomenon, coupled with helicity-dependent interband selection rules, allows for optical initialization and readout of a coupled spin-valley polarization~\cite{Xiaodong2014}.
Many combinations of different TMDC MLs yield a type-II band alignment, which leads to interlayer charge separation. Spatially separated electron-hole pairs can form so-called interlayer excitons (ILE) in these heterobilayers~\cite{Rivera2018}. Depending on the specific material combination, these ILE may be optically bright only for  specific crystallographic alignments~\cite{Nayak17} (interlayer twist) \blue{and} their energy may be tunable via control of interlayer twist~\cite{Kunstmann18,TEBYETEKERWA2021100509}. These ILE  inherit some properties, such as spin-valley polarization~\cite{Xu_Science16}, from the constituent TMDC MLs. However, in contrast to monolayer excitons, they are characterized by ultralong lifetimes~\cite{Xu_NatComm15,Wurstbauer17,Nagler17} and diffusion over mesoscopic distances~\cite{Xu_Science16,Unuchek2018}, which makes them attractive for exciton-based optoelectronic devices, see  Ref.~\onlinecite{Ciarrocchi2022} for a recent review.

Magneto-optical studies in high magnetic fields have been used very successfully  to elucidate properties of TMDC monolayers, such as exciton g factors~\cite{macneill2015breaking},  magnetic-field-induced valley polarization~\cite{Mitioglu2015}, exciton Bohr radii and masses~\cite{Stier2016,Goryca2019}, dark exciton~\cite{Zhang2017} and Rydberg exciton states~\cite{Stier18,Shi-Rydberg}, as well as the substructure of more complex quasiparticles like biexcitons~\cite{Nagler18,Barbone2018,Li2018}, see also Ref.~\onlinecite{doi:10.1063/5.0042683} for a recent review. 
More recently, ILE in TMDC heterobilayers have also been subjected to high magnetic fields, revealing a unique ability to engineer their effective g factor by changing the twist angle~\cite{Nagler17b}, which can obtain values far larger than those observed in TMDC ML excitons or change its sign~\cite{Ciarrocchi2019,Seyler2019}. 

For the specific material combination of WSe$_2$ and MoSe$_2$, optically bright ILE are only observable for interlayer twist angles close to 0 or 60 degrees~\cite{Nayak17}. These configurations are also referred to as R-type (0 degree) or H-type (60 degree) in accordance to the prevalent stacking polymorphs of TMDC multilayers. The optical selection rules for ILE in these structures depend on the local interlayer atomic registry~\cite{yu2017moire}, and therefore, the helicity of the emitted PL is not directly linked to ILE valley polarization, in contrast to TMDC monolayers. In heterobilayers, the interlayer atomic registry can vary spatially  due to two different effects:
\begin{itemize}
	\item the formation of a moiré lattice~\cite{Seyler2019,Tran2019}, which arises from an angular misalignment of the individual layers.
	\item atomic reconstruction~\cite{Rosenberger20,Weston20}, in which the individual layers are slightly distorted to yield domains with perfect interlayer atomic registry separated by domain walls. 
\end{itemize}
Both effects lead to  exciton localization, which can be used to study highly tunable manybody phases of excitons and individual charge carriers~\cite{yu2017moire,brotons2020spin,shabani2021deep,zhang2021van} .

In two independent  magneto-optical studies on ILE in H-type structures~\cite{Nagler17b,Delhomme_2020}, a peculiar enhancement of ILE valley polarization was found in magnetic fields of about 24~Tesla. In the latter study, this enhancement was associated with a coupling between ILE and chiral optical phonons~\cite{PhysRevB.90.115438,PhysRevLett.115.115502,he2020valley}. The strong electron-phonon and exciton-phonon interactions were also shown to limit the mobility~\cite{Kaasbjerg2012,Song2013,PhysRevB.87.115418,Jin2014}, lead to formation of polarons~\cite{Christiansen2017,doi:10.1063/1.5030678,doi:10.1063/1.5025907,PhysRevB.100.041301}, and produce phonon cascades~\cite{chow2017phonon,PhysRevB.98.035302,Brem2018,paradisanos2021efficient}.

Here, we present a joint experimental and theoretical study of ILE in both H-type and R-type  WSe$_2$-MoSe$_2$ heterobilayers. In magneto-photoluminescence measurements, we observe a pronounced enhancement of the ILE valley polarization at about 24.2~Tesla for both types of structure, even though their g factors, and the corresponding valley Zeeman splitting of the ILE, differ by about a factor of 3. This observation is explained as a valley-magnetophonon resonance of a hole in the localized exciton. Our theoretical analysis reveals the dominant mechanism of the valley-magnetophonon resonance to be electron-spin-conserving scattering with a chiral TA phonon originating from MoSe$_2$ ML between the excitonic states mixed by the long-range exchange interaction.

The magnetophonon resonance was predicted more than half a century ago by V. L. Gurevich and Yu. A. Firsov~\cite{gurevich1961theory} as an intrinsic resonance between a pair of Landau levels and the optical phonon energy. Soon after, the spin-magnetophonon resonance between opposite electron spin states was predicted~\cite{firsov1965effect} and observed~\cite{Akselrod1965}. Spin-conserving intervalley magnetophonon resonances were also studied in conventional semiconductors~\cite{FIRSOV19911181}, graphene~\cite{Basko_2016} and TMDC MLs~\cite{PhysRevB.100.041301}. Eventually the magnetophonon resonance evolved into a powerful tool to study both the phonon and electron properties of metals, semiconductors and semiconductor \blue{nanostructures~\cite{FIRSOV19911181,PhysRevB.38.13133,PhysRevLett.66.794,PhysRevB.53.16481}}. However, despite  numerous investigations and applications of the magnetophonon resonance, an intervalley spin-flip resonance was never observed before to the best of our knowledge. Thus the hole valley-magnetophonon resonance represents a novel aspect of this tool highly relevant for TMDC HS.
  
\stoptocentries
\section{Experimental results}
\starttocentries

\begin{figure}[h!]
	\begin{center}
		\includegraphics[width=1.0\linewidth]{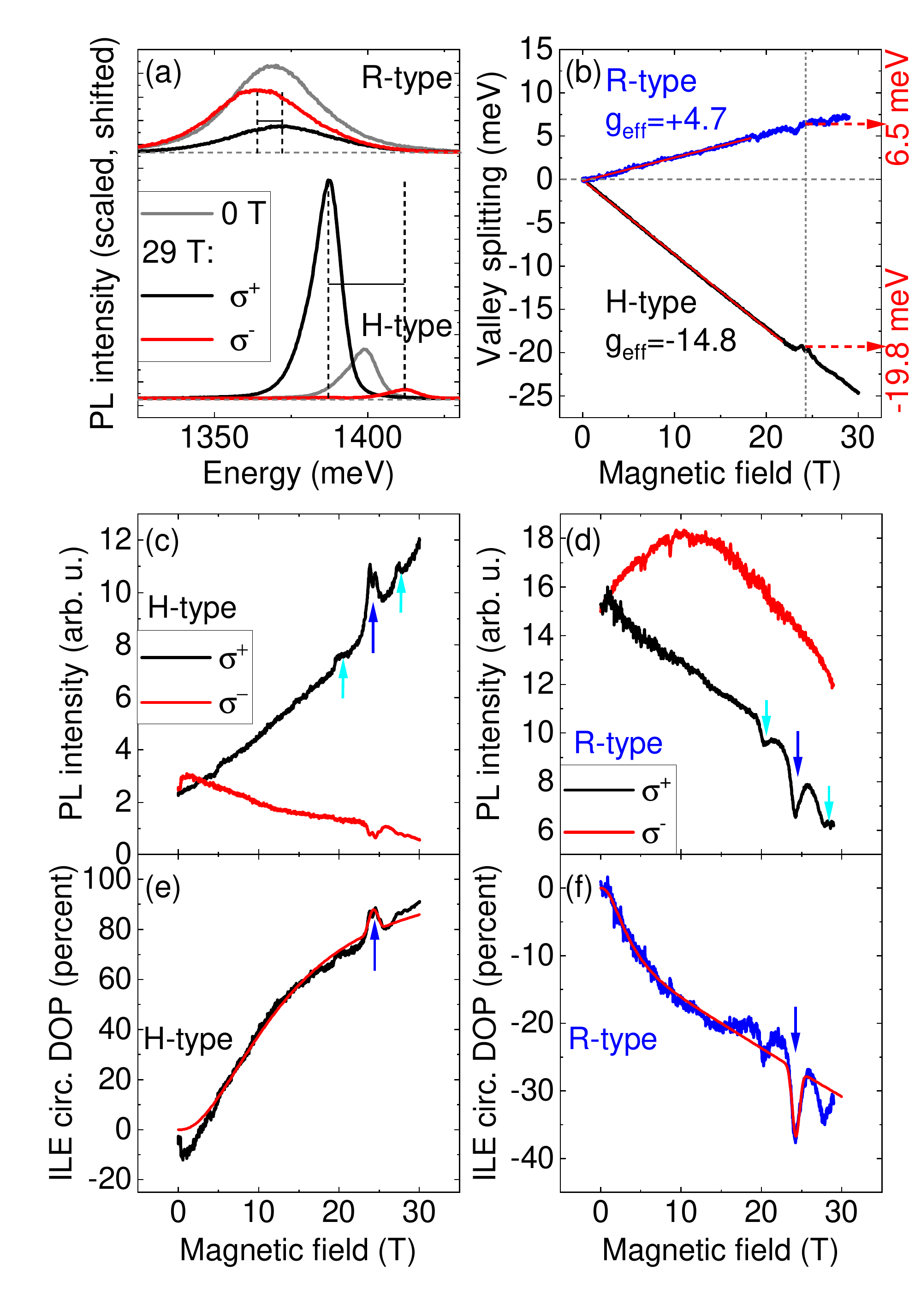}
	\end{center}
	\caption{(a) PL spectra of R-type and H-type HS at 0~T (grey lines) and helicity-resolved PL spectra at 29~T. The vertical dashed lines indicate the splitting between $\sigma^+$ and $\sigma^-$ emission at 29~T. (b) Valley splitting in  R-type (blue line) and H-type (black line) HS as a function of magnetic field. The red lines indicate linear fits to the data. The vertical dashed line indicates the resonance field of 24.2~T. The red arrows indicate the valley splitting values for the different HS at this field. (c+d) Helicity-resolved PL intensity for H-type (c) and R-type (d) HS as a function of magnetic field. The blue arrows indicate the resonantly increased (c) / decreased (d) PL intensity at the resonance field. The light blue arrows indicate the weaker resonant features. (e+f) ILE PL circular  degree of polarization for H-type (e) and R-type (f) HS calculated from data in (c+d). The red solid lines in (e) and (f) indicate fits to the data using our model. Blue arrows indicate the resonantly enhanced degree of polarization.}
	\label{Charakt-6Panel}
\end{figure}
Our  results are summarized in Fig.~\ref{Charakt-6Panel}. ILE in H-type and R-type heterobilayers are directly distinguishable by their emission energy and spectral linewidth, even in zero-field photoluminescence (PL) spectra~\cite{Holler22}. As Fig.~\ref{Charakt-6Panel}(a) shows, at 0~Tesla the ILE in the H-type structure has its emission peak at around 1399~meV, with a linewidth of about 13~meV. By contrast, the ILE in R-type structure emits at the significantly lower energy of about 1368~meV and has a larger linewidth of about 37~meV. In a high magnetic field (29~Tesla spectra are depicted in Fig.~\ref{Charakt-6Panel}(a)), helicity-resolved PL spectra reveal a pronounced energetic splitting of $\sigma^+$ and $\sigma^-$ polarized emission, combined with a pronounced change in relative emission intensities. For both  H-type and R-type ILE, the lower-energy emission becomes more intense than the high-energy emission. However, for the two structures, $\sigma^+$ and $\sigma^-$ components shift in opposite ways: for H-type, the $\sigma^-$ component shifts to higher energy, while for R-type, it is the   $\sigma^+$ component. It is also directly evident that the magnitude of the field-induced shifts is far larger in the H-type ILE. In order to quantify these observations, we performed continuous sweeps of the magnetic field from 0~T to 30~T (H-type) or 29~T(R-type), with helicity-resolved PL spectra taken at fixed time intervals corresponding to about 27~mT spacing between spectra. For each spectrum, the ILE signal was analyzed using an automatized Gaussian fit routine to extract its peak position and integrated intensity. From these datasets, we were able to determine the dependence of the valley splitting $\Delta E$ (defined as $\Delta E=E_{\sigma^+}-E_{\sigma^-})$ on magnetic field, as depicted in Fig.~\ref{Charakt-6Panel}(b). We clearly see a linear dependence for both types of ILE, with opposite sign and different slope. A linear fit yields the effective ILE g factors of $g_{eff}$=-14.8 for the H-type and $g_{eff}$=+4.7 for the R-type structure. Close to the resonance field of 24.2~T, we note a slight deviation of the measured valley splitting from the linear behavior for both structures. Noteworthy, the valley splitting at this resonance field differs by a factor of more than 3 between the structures, as indicated by the red arrows.

In addition to the valley-selective shifting of ILE energies, the magnetic field also modifies the relative intensities of $\sigma^+$ and $\sigma^-$ emission. We plot the helicity-resolved integrated PL intensities as a function of magnetic field for both structures in Fig.~\ref{Charakt-6Panel}(c) (H-type) and (d) (R-type), respectively. For the H-type ILE, the  $\sigma^+$ emission increases almost monotonously with magnetic field, while $\sigma^-$ decreases almost monotonously. However, we note a pronounced, resonant increase of the $\sigma^+$ emission at the resonance field of 24.2~T (marked by blue arrow). In the H-type structure, this is accompanied by a resonant reduction of the $\sigma^-$ emission at the same field. By contrast, in the R-type structure, the $\sigma^-$ initially increases up to about 10~T, then decreases. The $\sigma^+$ emission  decreases almost monotonously, but we notice a pronounced, resonant \textit{decrease} at the resonance field (marked by blue arrow). In the R-type structure, this is not accompanied by an increased emission in the opposite helicity. Looking more closely, we also see two weaker resonant features for both structures at fields slightly above and below the resonance field (marked by light blue arrows).

From these datasets, we calculate the circular degree of polarization (DOP) of the ILE emission, defined as
\begin{equation}\label{Dop}
 \text{DOP} = \frac{I^{\sigma^+}-I^{\sigma^-}}{I^{\sigma^+}+I^{\sigma^-}}  
\end{equation}
with the helicity-resolved PL intensities $I^{\sigma^+}$ and $I^{\sigma^-}$. The DOP as a function of magnetic field is depicted in Fig.~\ref{Charakt-6Panel}(e) and (f).  We note that, based on our definition, it is positive for the H-type structure and negative for the R-type structure. For both ILE types, the absolute value of the DOP increases as a function of magnetic field. In both cases, we clearly see a resonantly increased absolute value of the DOP at the resonance field, accompanied by two additional, weaker resonant features below and above the main resonance. While the DOP for the H-type structure reaches near-unity values above 90~percent at the largest applied magnetic field, the maximum absolute value for the R-type structure is lower at about 37~percent and actually achieved at the resonance field. This  difference in the maximum DOP closely corresponds to the difference of the valley splittings.

Our most surprising observation is the resonant enhancement of the DOP at the same field of 24.2~T despite the large difference of the valley splittings. Below, we demonstrate that this is a consequence of the hole valley-magnetophonon resonance.

\stoptocentries
\section{Theory}
\starttocentries

The effective exciton g factors for H-type and R-type HS and bright PL agree with the dominant contribution of H$_h^h$ (AA$'$)~\cite{Nagler17b,PhysRevB.100.041402,brotons2020spin} and R$_h^X$ (A$'$B$'$)~\cite{Ciarrocchi2019,electrically-controlled} interlayer atomic registries to the optical properties in agreement with the previous studies of MoSe$_2$/WSe$_2$ HS. These g factors stem from the individual electron and hole g factors as $-g_e\mp g_h$, respectively, where ``hole'' refers to the vacant state in the valence band. From the measured values of $-14.8$ and $+4.7$ we estimate the electron and hole g factors to be $g_e=5.05$ and $g_h=9.75$ in agreement with  first principle calculations~\cite{wozniak2020exciton,PhysRevResearch.2.033256,Deilmann2020,Forste2020a}.

The resonant changes of the PL at the same magnetic field $B_{\text{res}}=24.2$~T in both H-type and R-type HS suggest a common resonance. Despite the large difference in the exciton valley splittings the individual electron and hole Zeeman energies are the same at the given magnetic field for both stacking configurations. Therefore we attribute the observed resonances to the individual charge carriers.

Electron or hole intervalley scattering requires a spin flip and absorption or emission of a chiral phonon at the corner of the Brillouin zone (K points). The large density of chiral phonon states strongly increases the scattering rate. Due to the spin-valley locking the observed resonance represents a spin-valley magnetophonon resonance. Intervalley spin-magnetophonon resonances were never observed before to the best of our knowledge.

The electron and hole Zeeman splittings in the field $B_{\text{res}}$ are $g_e\mu_BB_{\text{res}}=7.1$~meV and $g_h\mu_BB_{\text{res}}=13.7$~meV. Calculations of the phonon energies in MoSe$_2$ and WSe$_2$~\cite{Song2013,Horzum2013,doi:10.1063/1.4794363,C5RA19747C,lin2021narrow,Lin21} demonstrate the absence of K phonon modes  at the Zeeman splitting of the electron. However in the vicinity of the hole Zeeman splitting there are the chiral ZA phonon mode of WSe$_2$ at $15$~meV and the chiral TA phonon mode of MoSe$_2$ at $14.7$~meV. Taking into account the possible phonon energy renormalization~\cite{https://doi.org/10.1002/pssb.202100321,Parzefall_2021} we conclude that we observe a valley-magnetophonon resonance of a hole.

\stoptocentries
\subsection{Valley magnetophonon resonance in PL polarization}
\starttocentries

\begin{figure}[h!]
	\begin{center}
		\includegraphics[width=1.0\linewidth]{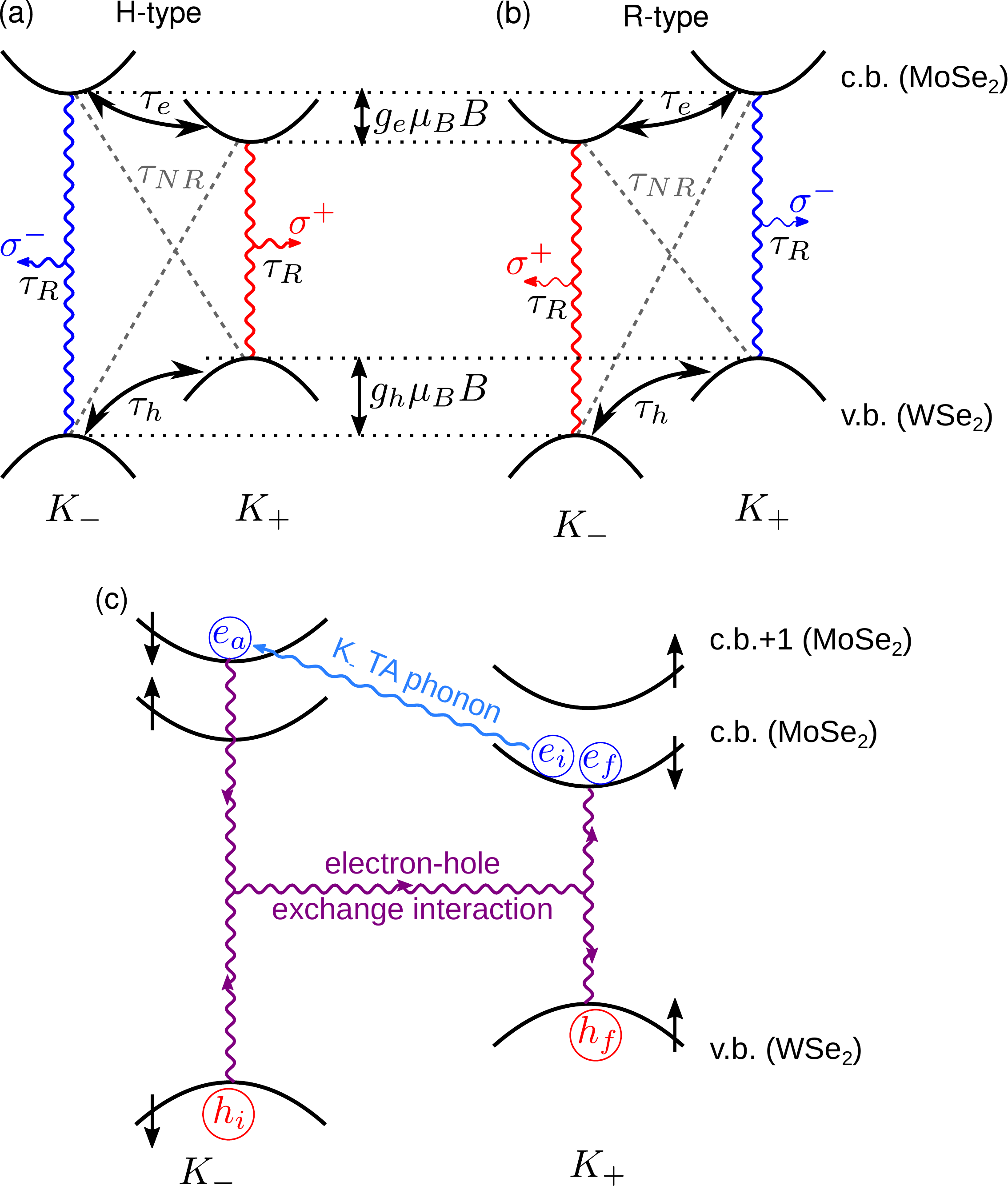}
	\end{center}
	\caption{(a,b) Optical selection rules in H$_h^h$ (a) and R$_h^X$ (b) interlayer atomic registries of MoSe$_2$/WSe$_2$ HS, electron ($\tau_e$) and hole ($\tau_h$) intervalley scatterings, radiative ($\tau_R$) and nonradiative ($\tau_{NR}$) electron hole recombinations, and Zeeman splittings of the conduction ($g_e\mu_BB$) and valence ($g_h\mu_BB$) bands. (c) Two step hole intervalley scattering involving electron-spin-conserving intervalley scattering with chiral TA phonon and exciton intervalley mixing by the long range exchange interaction. The electron and hole states are denoted as $e_i$, $h_i$ (initial), $e_a$ (auxiliary), and $e_f$, $h_f$ (final). The Zeeman splittings and spin states (black arrows) correspond to H-type HS.}
	\label{Theory-3Panel}
\end{figure}

In order to demonstrate that the hole valley magnetophonon resonance leads to the resonant enhancement of the PL polarization we consider the exciton spin dynamics using  rate equations for the four lowest intravalley and intervalley excitonic states shown in Fig.~\ref{Theory-3Panel}(a,b)~\cite{SI}. We take into account radiative and nonradiative exciton recombination times, $\tau_R$ and $\tau_{NR}$, respectively, as well as the electron, $\tau_e(B)$, and hole, $\tau_h(B)$, intervalley scattering times to the lower energy states. \blue{Note that the actual hole scattering mechanism may be quite complex and involve both charge carriers in a single scattering event, as shown in the next subsection.} The scattering times to the higher energy states are larger by the corresponding factors $\exp(g_{e,h}\mu_BB/k_B T)$ with $T$ being the temperature and $k_B$ being the Boltzmann constant.

For the hole valley relaxation time we assume the following form:
\begin{equation}
  \frac{1}{\tau_h(B)}=\frac{1}{\tau_{\text{res}}}\exp\left[-\frac{(B-B_{\text{res}})^2}{\Delta B^2}\right]+\frac{1}{\tau_h^{(0)}}\left(\frac{B}{B_{\text{res}}}\right)^2.
\end{equation}
Here the first term stands for the resonant intervalley scattering at field $B_{\text{res}}$ with the minimum time $\tau_{\text{res}}$ and the width of the resonance $\Delta B$. The second term describes the phenomenological scattering time $\tau_h(B)\propto1/B^2$, which corresponds to the direct spin-phonon coupling in strained HS in small magnetic fields~\cite{Pearce2017}. Similarly to this, for the electron intervalley scattering we assume $\tau_e(B)=\tau_e^{(0)}(B_{\text{res}}/B)^2$.

Figure~\ref{Charakt-6Panel}(e,f) shows that this model nicely fits the polarization degree of PL almost in the whole range of magnetic fields from 0 to 29~T including the resonant enhancement at 24.2~T. The fit parameters are given and discussed in the Supplementary material~\cite{SI}. The surprisingly good quality of the fit demonstrates that the resonant enhancement of the PL polarization degree is related to the valley-magnetophonon resonance.

\stoptocentries
\subsection{Mechanism of hole valley-magnetophonon resonance}
\starttocentries

Direct intervalley scattering between Kramers degenerate states of the hole is forbidden by the time reversal symmetry~\cite{birpikus_eng,ivchenko1990current,Song2013}. However, it becomes possible in the presence of the external magnetic field~\cite{PhysRevB.64.125316} or hole-electron exchange interaction~\cite{PhysRevB.67.205330}. The excitons in our MoSe$_2$/WSe$_2$ HS are localized either  due to  the moir\'e potential or the domains formed by atomic reconstruction. We find that the dominant mechanism of the hole valley-magnetophonon resonance in this case is a two step process~\cite{SI}, as illustrated in Fig.~\ref{Theory-3Panel}(c). To be specific, \blue{let us consider the scattering from an intervalley excitonic state with longer lifetime to an intravalley state with a short lifetime with the transition of a hole in exciton to a state with lower energy at the $K_+$ valley, as shown in Fig.~\ref{Theory-3Panel}(c).} At the first step, an \emph{electron} from the $K_+$ valley virtually scatters to the upper (spin split) subband in the $K_-$ valley emitting a chiral TA phonon of the MoSe$_2$ ML. \blue{The phonon emission ensures the energy conservation for the entire two-step scattering, and the intermediate (auxiliary) exciton state has a very short lifetime limited by the time-energy uncertainty relation.} At the second step the \textit{exciton} scatters as a whole from the $K_-$ valley to the ground state in the $K_+$ valley due to the long-range electron hole exchange interaction. \blue{This step can be described as emission and reabsorption of a virtual longitudinal photon~\cite{birpikus_eng,Goupalov03_eng,Glazov2014}. The efficiency of this scattering is ensured by the brightened spin-triplet exciton states and specific optical selection rules for H$_h^h$ and R$_h^X$ interlayer atomic registries~\cite{Yu_2018}. In total,} the electron remains in the same state, and the hole flips its valley. The hole scattering from intravalley exciton state in $K_-$ to $K_+$ valley has the same rate, and the \blue{scattering with increase of the energy} is suppressed by the Boltzmann factor.

The scattering rate is given by the Fermi's golden rule
\begin{equation}
  \frac{1}{\tau_{\text{res}}}=\frac{2\pi}{\hbar}\sum_{\bm q}\left|\frac{\left<f_{\bm q}\middle|\mathcal H_{\text{exch}}\middle|a_{\bm q}\right>\left<a_{\bm q}\middle|\mathcal H_{\text{e-ph}}\middle|i\right>}{E_f-E_a}\right|^2\delta(E_i-E_f-\hbar\Omega_{q}),
\end{equation}
where $i$, $a_{\bm q}$, $f_{\bm q}$ represent the initial, auxiliary and final states of exciton and emitted phonon with the wave vector $\bm K_-+\bm q$ and energy $\hbar\Omega_{q}$, $E_{i,a,f}$ are the exciton energies in the corresponding states, and $\mathcal H_{\text{exch}}$ and $\mathcal H_{\text{e-ph}}$ stand for the exchange and electron-phonon interaction Hamiltonians. In the vicinity of the magnetophonon resonance one has $E_a-E_f=\Delta_c+(g_h-g_e^v)\mu_B B_{\text{res}}$, where $\Delta_c$ is the spin orbit splitting of MoSe$_2$ conduction band and $g_e^v$ is the electron valley g factor~\cite{Wang2015b,DurnevUFN}.

From the symmetry analysis we find that the electron interaction with chiral TA phonons in MoSe$_2$ ML is described by~\cite{SI}
\begin{equation}
  \mathcal H_{\text{e-ph}}=\sum_{\bm q,\pm}\sqrt{\frac{\hbar}{2\rho\Omega_qA}}\Xi q_\pm\blue{\hat{\tau}_\pm}b_{\bm K_\pm+\bm q}^\dag\e^{-\i\bm q\bm r_e}+{\text{H.c.}},
\end{equation}
where \blue{$b_{\bm k}^\dag$} are the phonon creation operators, $\rho$ is \blue{two-dimensional mass density} of the ML, $A$ is the normalization area, $q_\pm=(q_x\pm\i q_y)/\sqrt{2}$, $\Xi$ is the intervalley deformation potential, $\bm r_e$ is the electron coordinate, and $\blue{\hat{\tau}_\pm}$ are the electron valley rising and lowering operators, which conserve the electron spin. This Hamiltonian can be derived taking into account that electrons and phonons at $K_\pm$ valleys have orbital angular momenta $\pm 1$ and $\mp 1$, respectively, and that the angular momentum modulo 3 should be conserved during the scattering. The phonon dispersion at the corners of the Brillouin zone has the form $\hbar\Omega_q=\hbar\Omega_0+(\hbar q)^2/(2M)$~\cite{C5RA19747C}, where $M$ is the effective phonon mass at $K$ point.

The Hamiltonian of the long-range exchange interaction between auxiliary and final states can be obtained similarly to the ML case~\cite{Glazov2014}, it reads
\begin{equation}
  \mathcal H_{\text{exch}}=\frac{2\pi e^2\delta(\bm\rho)}{\varkappa_bm_0^2\omega_0^2}\frac{(\bm K\bm p_{\text{c.v.}}^a)(\bm K\bm p_{\text{c.v.}}^f)^*}{K},
\end{equation}
where $\bm\rho$  is the distance between electron and hole, $\varkappa_b$ is the background dielectric constant, $m_0$ is the free electron mass, $\omega_0$ is the exciton resonance frequency, $\bm K$ is the exciton center of mass momentum, and $p_{c.v.}^{a,f}$ are the interband momentum matrix elements for the auxiliary and final states.

To calculate the scattering rate we consider the wave function of the localized exciton $\propto\e^{-\rho/a_B-R^2/(2l^2)}$, where $a_B$ and $l$ are the exciton Bohr radius and the localization length, and $\bm R$ is the exciton center of mass coordinate. Then under assumption of the low temperature, $k_BT\ll g_h\mu_BB_{\text{res}}$ we obtain the hole intervalley scattering rate~\cite{SI}
\begin{multline}
  \frac{1}{\tau_{\text{res}}}=\frac{\pi mM^2\Gamma_0^a\Gamma_0^fE_{\text{loc}}\Xi^2(B-B_{\text{res}})}{4\hbar^3\rho k^2\left(E_a-E_f\right)^2B_{\text{res}}}\theta(B-B_{\text{res}})\\
  \times\exp\left(-\frac{Mg_h\mu_B(B-B_{\text{res}})}{mE_{\text{loc}}}\right),
\end{multline}
where $E_{\text{loc}}=\hbar^2/(ml^2)$ is the exciton localization energy with $m$ being the exciton mass and $\Gamma_0^{a,f}=2\pi k^2e^2|p_{\text{c.v.}}^{a,f}|^2|\varphi(0)|^2/(\hbar\varkappa_b\omega_0^2m_0^2)$ are the free exciton radiative decay rates in the auxiliary and final states with $\varphi(0)=\sqrt{2/\pi}/a_B$ being the wave function of the relative electron hole motion at $\rho=0$ and $k=\sqrt{\varkappa_b}\omega_0/c$ being the light wave vector.

One can see that the hole intervalley scattering rate vanishes at magnetic fields below $B_{\text{res}}$, as there are no phonons of the required energy, which is described by the Heaviside step function $\theta(B-B_{\text{res}})$. Just above $B_{\text{res}}$ the scattering rate grows linearly with increase of $B-B_{\text{res}}$ because of the increase of the electron-phonon interaction matrix element. However at high magnetic fields the exciton-phonon matrix element decreases exponentially because of the exciton localization. As a result, the hole intervalley scattering rate has a narrow maximum at $B\approx B_{\text{res}}$ with the width of the order of $mE_{\text{loc}}/(g_h\mu_BM)$. In the maximum it reaches
\begin{equation}
  \frac{1}{\tau_{\text{res}}}=\frac{\pi m^2M\Gamma_0^a\Gamma_0^fE_{\text{loc}}^2\Xi^2}{4\hbar^3\rho k^2(E_a-E_f)^2g_h\mu_BB_{\text{res}}}.
\end{equation}
Substitution of the material parameters yields~\cite{SI} $\tau_{\text{res}}=12$~ns and $20$~ns for H-type and R-type HS, which agrees with the timescales of the PL polarization saturation~\cite{Holler22}.

\stoptocentries
\section{Discussion and conclusion}
\starttocentries

The resonant increase of the PL polarization in the same magnetic field for H-type and R-type HS unambiguously reveals a valley-magnetophonon resonance. The weaker resonance below and higher by approximately 3.8~T for both HS may be related to some combined phonon resonances, but their exact nature is not completely clear.

The dominant microscopic mechanism of the hole valley-magnetophonon resonance is found to be the scattering with chiral TA phonon of MoSe$_2$ between the excitonic states mixed by the long-range exchange interaction. Noteworthy it has a few solid advantages: (i) it does not require spin-dependent electron-phonon interaction, (ii) it profits from the large density of chiral phonon states, (iii) it is free of the Van Vleck cancellation, (iv) the long-range exchange interaction is enhanced by the exciton localization, and (v) optical selection rules for excitons in H$_h^h$ and $R_h^X$ atomic registries exactly match the requirements for the exchange interaction. The unique observation of the valley-magnetophonon resonance in TMDC HS is made possible by the strong spin splittings of the bands.

In summary, we have observed a hole valley-magnetophonon resonance of interlayer excitons localized at reconstructed/moir\'e potential in both H-type and R-type MoSe$_2$/WSe$_2$ HS at magnetic field of 24.2~T. It leads to the resonant enhancement of the PL polarization degree under nonresonant excitation at low temperatures. The hole intervalley scattering involves a chiral TA phonon originating from the MoSe$_2$ ML and long-range exciton exchange interaction. The valley-magnetophonon resonance is important for both the transport properties in moir\'e HS and optical manipulation of the valley degree of freedom of charge carriers.

\stoptocentries
\acknowledgements 
\starttocentries

We thank M. M. Glazov and S. A. Tarasenko for fruitful discussions, B. Peng and K. Lin for sharing the phonon dispersion curves, RF President Grant No. MK-5158.2021.1.2, the Foundation for the Advancement of Theoretical Physics and Mathematics ``BASIS''. We gratefully acknowledge financial support by the DFG via the following projects: GRK 1570 (J. H., P.N., C.S.), KO3612/3-1(project-ID 631210, T.K.), KO3612/4-1(project-ID 648265, T.K.), SFB1277 (project B05, T.K., M.K., A.C., C.S.), SFB1477 (project-ID 441234705, T.K.) Emmy-Noether Programme (CH 1672/1, A.C.), W\"urzburg-Dresden Cluster of Excellence on Complexity and Topology in Quantum Matter ct.qmat (EXC 2147, Project-ID 390858490, A.C.),  Walter-Benjamin Programme (project-ID 462503440, J.Z.), and SPP2244 (project-ID 670977 C.S., T.K.).
This work was supported by HFML-RU/NWO-I, member of the European Magnetic Field Laboratory (EMFL).
The theoretical analysis of the mechanism of the hole valley-magnetophonon resonance by D.S.S. was supported by the \blue{Russian Foundation for Basic Research} Grant No. 19-52-12038.

\stoptocentries
\section{Methods}
\starttocentries

\stoptocentries
\subsection{Sample preparation}
\starttocentries

Our heterostructures  were fabricated by means of a deterministic transfer process~\cite{Castellanos2014} using  bulk crystals supplied by HQ graphene. Monolayers of the constituent materials are prepared on an intermediate polydimethylsiloxane substrate and subsequently stacked on top of each other on a silicon substrate covered with a silicon oxide layer. In order to achieve crystallographic alignment, well-cleaved edges of the constituent layers are aligned parallel to each other during the transfer. Further details are published elsewhere~\cite{Holler22}. 

\stoptocentries
\subsection{Optical spectroscopy}
\starttocentries

Low-temperature PL measurements in high magnetic fields were performed at the HFML facility in Nijmegen. The sample was placed on a x-y-z piezoelectric stage and cooled down to 4.2~K in a cryostat filled with liquid helium. Magnetic fields up to 30~T were applied by means of a resistive magnet in Faraday configuration. 
A diode laser (emission wavelength 640~nm) was used for excitation. The laser light was
linearly polarized and focused onto the sample with a microscope objective resulting in a spot size of about 4~$\mu$m. The polarization of the PL was analyzed with a quarter-wave plate and a linear polarizer. The PL was then coupled into a grating spectrometer, where it was detected using a CCD sensor. 
For  field sweeps, the magnetic field was ramped continuously from 0~T to up to 30~T (for technical reasons, the field sweep  for the R-type HS was limited to 29~T), and spectra for a fixed detection helicity were recorded at fixed time intervals. At the maximum field, the detection helicity was flipped and the field was ramped down continuously to 0~T, so that spectra for the other helicity could be recorded.

\renewcommand{\i}{\ifr}
\let\oldaddcontentsline\addcontentsline
\renewcommand{\addcontentsline}[3]{}

\begin{thebibliography}{86}
\expandafter\ifx\csname natexlab\endcsname\relax\def\natexlab#1{#1}\fi
\expandafter\ifx\csname bibnamefont\endcsname\relax
  \def\bibnamefont#1{#1}\fi
\expandafter\ifx\csname bibfnamefont\endcsname\relax
  \def\bibfnamefont#1{#1}\fi
\expandafter\ifx\csname citenamefont\endcsname\relax
  \def\citenamefont#1{#1}\fi
\expandafter\ifx\csname url\endcsname\relax
  \def\url#1{\texttt{#1}}\fi
\expandafter\ifx\csname urlprefix\endcsname\relax\def\urlprefix{URL }\fi
\providecommand{\bibinfo}[2]{#2}
\providecommand{\eprint}[2][]{\url{#2}}

\bibitem[{\citenamefont{Mak et~al.}(2010)\citenamefont{Mak, Lee, Hone, Shan,
  and Heinz}}]{Mak2010}
\bibinfo{author}{\bibfnamefont{K.~F.} \bibnamefont{Mak}},
  \bibinfo{author}{\bibfnamefont{C.}~\bibnamefont{Lee}},
  \bibinfo{author}{\bibfnamefont{J.}~\bibnamefont{Hone}},
  \bibinfo{author}{\bibfnamefont{J.}~\bibnamefont{Shan}}, \bibnamefont{and}
  \bibinfo{author}{\bibfnamefont{T.~F.} \bibnamefont{Heinz}},
  \bibinfo{journal}{Phys. Rev. Lett.} \textbf{\bibinfo{volume}{105}},
  \bibinfo{pages}{136805} (\bibinfo{year}{2010}).

\bibitem[{\citenamefont{Splendiani et~al.}(2010)\citenamefont{Splendiani, Sun,
  Zhang, Li, Kim, Chim, Galli, and Wang}}]{Splendiani2010}
\bibinfo{author}{\bibfnamefont{A.}~\bibnamefont{Splendiani}},
  \bibinfo{author}{\bibfnamefont{L.}~\bibnamefont{Sun}},
  \bibinfo{author}{\bibfnamefont{Y.}~\bibnamefont{Zhang}},
  \bibinfo{author}{\bibfnamefont{T.}~\bibnamefont{Li}},
  \bibinfo{author}{\bibfnamefont{J.}~\bibnamefont{Kim}},
  \bibinfo{author}{\bibfnamefont{C.-Y.} \bibnamefont{Chim}},
  \bibinfo{author}{\bibfnamefont{G.}~\bibnamefont{Galli}}, \bibnamefont{and}
  \bibinfo{author}{\bibfnamefont{F.}~\bibnamefont{Wang}},
  \bibinfo{journal}{Nano Letters} \textbf{\bibinfo{volume}{10}},
  \bibinfo{pages}{1271} (\bibinfo{year}{2010}).

\bibitem[{\citenamefont{Chernikov et~al.}(2014)\citenamefont{Chernikov,
  Berkelbach, Hill, Rigosi, Li, Aslan, Reichman, Hybertsen, and
  Heinz}}]{Chernikov2014}
\bibinfo{author}{\bibfnamefont{A.}~\bibnamefont{Chernikov}},
  \bibinfo{author}{\bibfnamefont{T.~C.} \bibnamefont{Berkelbach}},
  \bibinfo{author}{\bibfnamefont{H.~M.} \bibnamefont{Hill}},
  \bibinfo{author}{\bibfnamefont{A.}~\bibnamefont{Rigosi}},
  \bibinfo{author}{\bibfnamefont{Y.}~\bibnamefont{Li}},
  \bibinfo{author}{\bibfnamefont{O.~B.} \bibnamefont{Aslan}},
  \bibinfo{author}{\bibfnamefont{D.~R.} \bibnamefont{Reichman}},
  \bibinfo{author}{\bibfnamefont{M.~S.} \bibnamefont{Hybertsen}},
  \bibnamefont{and} \bibinfo{author}{\bibfnamefont{T.~F.} \bibnamefont{Heinz}},
  \bibinfo{journal}{Phys. Rev. Lett.} \textbf{\bibinfo{volume}{113}},
  \bibinfo{pages}{076802} (\bibinfo{year}{2014}).

\bibitem[{\citenamefont{Xiao et~al.}(2012)\citenamefont{Xiao, Liu, Feng, Xu,
  and Yao}}]{Xiao2012}
\bibinfo{author}{\bibfnamefont{D.}~\bibnamefont{Xiao}},
  \bibinfo{author}{\bibfnamefont{G.-B.} \bibnamefont{Liu}},
  \bibinfo{author}{\bibfnamefont{W.}~\bibnamefont{Feng}},
  \bibinfo{author}{\bibfnamefont{X.}~\bibnamefont{Xu}}, \bibnamefont{and}
  \bibinfo{author}{\bibfnamefont{W.}~\bibnamefont{Yao}},
  \bibinfo{journal}{Phys. Rev. Lett.} \textbf{\bibinfo{volume}{108}},
  \bibinfo{pages}{196802} (\bibinfo{year}{2012}).

\bibitem[{\citenamefont{Xu et~al.}(2014)\citenamefont{Xu, Yao, Xiao, and
  Heinz}}]{Xiaodong2014}
\bibinfo{author}{\bibfnamefont{X.}~\bibnamefont{Xu}},
  \bibinfo{author}{\bibfnamefont{W.}~\bibnamefont{Yao}},
  \bibinfo{author}{\bibfnamefont{D.}~\bibnamefont{Xiao}}, \bibnamefont{and}
  \bibinfo{author}{\bibfnamefont{T.~F.} \bibnamefont{Heinz}},
  \bibinfo{journal}{Nat Phys} \textbf{\bibinfo{volume}{10}},
  \bibinfo{pages}{343} (\bibinfo{year}{2014}), ISSN \bibinfo{issn}{1745-2473}.

\bibitem[{\citenamefont{Rivera et~al.}(2018)\citenamefont{Rivera, Yu, Seyler,
  Wilson, Yao, and Xu}}]{Rivera2018}
\bibinfo{author}{\bibfnamefont{P.}~\bibnamefont{Rivera}},
  \bibinfo{author}{\bibfnamefont{H.}~\bibnamefont{Yu}},
  \bibinfo{author}{\bibfnamefont{K.~L.} \bibnamefont{Seyler}},
  \bibinfo{author}{\bibfnamefont{N.~P.} \bibnamefont{Wilson}},
  \bibinfo{author}{\bibfnamefont{W.}~\bibnamefont{Yao}}, \bibnamefont{and}
  \bibinfo{author}{\bibfnamefont{X.}~\bibnamefont{Xu}},
  \bibinfo{journal}{Nature Nanotechnology} \textbf{\bibinfo{volume}{13}},
  \bibinfo{pages}{1004} (\bibinfo{year}{2018}).

\bibitem[{\citenamefont{Nayak et~al.}(2017)\citenamefont{Nayak, Horbatenko,
  Ahn, Kim, Lee, Ma, Jang, Lim, Kim, Ryu et~al.}}]{Nayak17}
\bibinfo{author}{\bibfnamefont{P.~K.} \bibnamefont{Nayak}},
  \bibinfo{author}{\bibfnamefont{Y.}~\bibnamefont{Horbatenko}},
  \bibinfo{author}{\bibfnamefont{S.}~\bibnamefont{Ahn}},
  \bibinfo{author}{\bibfnamefont{G.}~\bibnamefont{Kim}},
  \bibinfo{author}{\bibfnamefont{J.-U.} \bibnamefont{Lee}},
  \bibinfo{author}{\bibfnamefont{K.~Y.} \bibnamefont{Ma}},
  \bibinfo{author}{\bibfnamefont{A.-R.} \bibnamefont{Jang}},
  \bibinfo{author}{\bibfnamefont{H.}~\bibnamefont{Lim}},
  \bibinfo{author}{\bibfnamefont{D.}~\bibnamefont{Kim}},
  \bibinfo{author}{\bibfnamefont{S.}~\bibnamefont{Ryu}}, \bibnamefont{et~al.},
  \bibinfo{journal}{ACS Nano} \textbf{\bibinfo{volume}{11}},
  \bibinfo{pages}{4041} (\bibinfo{year}{2017}).

\bibitem[{\citenamefont{Kunstmann et~al.}(2018)\citenamefont{Kunstmann,
  Mooshammer, Nagler, Chaves, Stein, Paradiso, Plechinger, Strunk,
  Sch\"{u}ller, Seifert et~al.}}]{Kunstmann18}
\bibinfo{author}{\bibfnamefont{J.}~\bibnamefont{Kunstmann}},
  \bibinfo{author}{\bibfnamefont{F.}~\bibnamefont{Mooshammer}},
  \bibinfo{author}{\bibfnamefont{P.}~\bibnamefont{Nagler}},
  \bibinfo{author}{\bibfnamefont{A.}~\bibnamefont{Chaves}},
  \bibinfo{author}{\bibfnamefont{F.}~\bibnamefont{Stein}},
  \bibinfo{author}{\bibfnamefont{N.}~\bibnamefont{Paradiso}},
  \bibinfo{author}{\bibfnamefont{G.}~\bibnamefont{Plechinger}},
  \bibinfo{author}{\bibfnamefont{C.}~\bibnamefont{Strunk}},
  \bibinfo{author}{\bibfnamefont{C.}~\bibnamefont{Sch\"{u}ller}},
  \bibinfo{author}{\bibfnamefont{G.}~\bibnamefont{Seifert}},
  \bibnamefont{et~al.}, \bibinfo{journal}{Nat. Phys.}
  \textbf{\bibinfo{volume}{14}}, \bibinfo{pages}{801} (\bibinfo{year}{2018}).

\bibitem[{\citenamefont{Tebyetekerwa et~al.}(2021)\citenamefont{Tebyetekerwa,
  Zhang, Saji, Wibowo, Rahman, Truong, Lu, Yin, Macdonald, and
  Nguyen}}]{TEBYETEKERWA2021100509}
\bibinfo{author}{\bibfnamefont{M.}~\bibnamefont{Tebyetekerwa}},
  \bibinfo{author}{\bibfnamefont{J.}~\bibnamefont{Zhang}},
  \bibinfo{author}{\bibfnamefont{S.~E.} \bibnamefont{Saji}},
  \bibinfo{author}{\bibfnamefont{A.~A.} \bibnamefont{Wibowo}},
  \bibinfo{author}{\bibfnamefont{S.}~\bibnamefont{Rahman}},
  \bibinfo{author}{\bibfnamefont{T.~N.} \bibnamefont{Truong}},
  \bibinfo{author}{\bibfnamefont{Y.}~\bibnamefont{Lu}},
  \bibinfo{author}{\bibfnamefont{Z.}~\bibnamefont{Yin}},
  \bibinfo{author}{\bibfnamefont{D.}~\bibnamefont{Macdonald}},
  \bibnamefont{and} \bibinfo{author}{\bibfnamefont{H.~T.}
  \bibnamefont{Nguyen}}, \bibinfo{journal}{Cell Reports Physical Science}
  \textbf{\bibinfo{volume}{2}}, \bibinfo{pages}{100509} (\bibinfo{year}{2021}).

\bibitem[{\citenamefont{Rivera et~al.}(2016)\citenamefont{Rivera, Seyler, Yu,
  Schaibley, Yan, Mandrus, Yao, and Xu}}]{Xu_Science16}
\bibinfo{author}{\bibfnamefont{P.}~\bibnamefont{Rivera}},
  \bibinfo{author}{\bibfnamefont{K.~L.} \bibnamefont{Seyler}},
  \bibinfo{author}{\bibfnamefont{H.}~\bibnamefont{Yu}},
  \bibinfo{author}{\bibfnamefont{J.~R.} \bibnamefont{Schaibley}},
  \bibinfo{author}{\bibfnamefont{J.}~\bibnamefont{Yan}},
  \bibinfo{author}{\bibfnamefont{D.~G.} \bibnamefont{Mandrus}},
  \bibinfo{author}{\bibfnamefont{W.}~\bibnamefont{Yao}}, \bibnamefont{and}
  \bibinfo{author}{\bibfnamefont{X.}~\bibnamefont{Xu}},
  \bibinfo{journal}{Science} \textbf{\bibinfo{volume}{351}},
  \bibinfo{pages}{688} (\bibinfo{year}{2016}).

\bibitem[{\citenamefont{Rivera et~al.}(2015)\citenamefont{Rivera, Schaibley,
  Jones, Ross, Wu, Aivazian, Klement, Seyler, Clark, Ghimire
  et~al.}}]{Xu_NatComm15}
\bibinfo{author}{\bibfnamefont{P.}~\bibnamefont{Rivera}},
  \bibinfo{author}{\bibfnamefont{J.~R.} \bibnamefont{Schaibley}},
  \bibinfo{author}{\bibfnamefont{A.~M.} \bibnamefont{Jones}},
  \bibinfo{author}{\bibfnamefont{J.~S.} \bibnamefont{Ross}},
  \bibinfo{author}{\bibfnamefont{S.}~\bibnamefont{Wu}},
  \bibinfo{author}{\bibfnamefont{G.}~\bibnamefont{Aivazian}},
  \bibinfo{author}{\bibfnamefont{P.}~\bibnamefont{Klement}},
  \bibinfo{author}{\bibfnamefont{K.}~\bibnamefont{Seyler}},
  \bibinfo{author}{\bibfnamefont{G.}~\bibnamefont{Clark}},
  \bibinfo{author}{\bibfnamefont{N.~J.} \bibnamefont{Ghimire}},
  \bibnamefont{et~al.}, \bibinfo{journal}{Nat. Commun.}
  \textbf{\bibinfo{volume}{6}}, \bibinfo{pages}{7242} (\bibinfo{year}{2015}).

\bibitem[{\citenamefont{Miller et~al.}(2017)\citenamefont{Miller, Steinhoff,
  Pano, Klein, Jahnke, Holleitner, and Wurstbauer}}]{Wurstbauer17}
\bibinfo{author}{\bibfnamefont{B.}~\bibnamefont{Miller}},
  \bibinfo{author}{\bibfnamefont{A.}~\bibnamefont{Steinhoff}},
  \bibinfo{author}{\bibfnamefont{B.}~\bibnamefont{Pano}},
  \bibinfo{author}{\bibfnamefont{J.}~\bibnamefont{Klein}},
  \bibinfo{author}{\bibfnamefont{F.}~\bibnamefont{Jahnke}},
  \bibinfo{author}{\bibfnamefont{A.}~\bibnamefont{Holleitner}},
  \bibnamefont{and}
  \bibinfo{author}{\bibfnamefont{U.}~\bibnamefont{Wurstbauer}},
  \bibinfo{journal}{Nano Lett.} \textbf{\bibinfo{volume}{17}},
  \bibinfo{pages}{5229} (\bibinfo{year}{2017}).

\bibitem[{\citenamefont{Nagler et~al.}(2017{\natexlab{a}})\citenamefont{Nagler,
  Plechinger, Ballottin, Mitioglu, Meier, Paradiso, Strunk, Chernikov,
  Christianen, Sch\"{u}ller et~al.}}]{Nagler17}
\bibinfo{author}{\bibfnamefont{P.}~\bibnamefont{Nagler}},
  \bibinfo{author}{\bibfnamefont{G.}~\bibnamefont{Plechinger}},
  \bibinfo{author}{\bibfnamefont{M.~V.} \bibnamefont{Ballottin}},
  \bibinfo{author}{\bibfnamefont{A.}~\bibnamefont{Mitioglu}},
  \bibinfo{author}{\bibfnamefont{S.}~\bibnamefont{Meier}},
  \bibinfo{author}{\bibfnamefont{N.}~\bibnamefont{Paradiso}},
  \bibinfo{author}{\bibfnamefont{C.}~\bibnamefont{Strunk}},
  \bibinfo{author}{\bibfnamefont{A.}~\bibnamefont{Chernikov}},
  \bibinfo{author}{\bibfnamefont{P.~C.~M.} \bibnamefont{Christianen}},
  \bibinfo{author}{\bibfnamefont{C.}~\bibnamefont{Sch\"{u}ller}},
  \bibnamefont{et~al.}, \bibinfo{journal}{2D Mater.}
  \textbf{\bibinfo{volume}{4}}, \bibinfo{pages}{025112}
  (\bibinfo{year}{2017}{\natexlab{a}}).

\bibitem[{\citenamefont{Unuchek et~al.}(2018)\citenamefont{Unuchek, Ciarrocchi,
  Avsar, Watanabe, Taniguchi, and Kis}}]{Unuchek2018}
\bibinfo{author}{\bibfnamefont{D.}~\bibnamefont{Unuchek}},
  \bibinfo{author}{\bibfnamefont{A.}~\bibnamefont{Ciarrocchi}},
  \bibinfo{author}{\bibfnamefont{A.}~\bibnamefont{Avsar}},
  \bibinfo{author}{\bibfnamefont{K.}~\bibnamefont{Watanabe}},
  \bibinfo{author}{\bibfnamefont{T.}~\bibnamefont{Taniguchi}},
  \bibnamefont{and} \bibinfo{author}{\bibfnamefont{A.}~\bibnamefont{Kis}},
  \bibinfo{journal}{Nature} \textbf{\bibinfo{volume}{560}},
  \bibinfo{pages}{340} (\bibinfo{year}{2018}).

\bibitem[{\citenamefont{Ciarrocchi et~al.}(2022)\citenamefont{Ciarrocchi,
  Tagarelli, Avsar, and Kis}}]{Ciarrocchi2022}
\bibinfo{author}{\bibfnamefont{A.}~\bibnamefont{Ciarrocchi}},
  \bibinfo{author}{\bibfnamefont{F.}~\bibnamefont{Tagarelli}},
  \bibinfo{author}{\bibfnamefont{A.}~\bibnamefont{Avsar}}, \bibnamefont{and}
  \bibinfo{author}{\bibfnamefont{A.}~\bibnamefont{Kis}},
  \bibinfo{journal}{Nature Reviews Materials}  (\bibinfo{year}{2022}).

\bibitem[{\citenamefont{MacNeill et~al.}(2015)\citenamefont{MacNeill, Heikes,
  Mak, Anderson, Korm{\'a}nyos, Z{\'o}lyomi, Park, and
  Ralph}}]{macneill2015breaking}
\bibinfo{author}{\bibfnamefont{D.}~\bibnamefont{MacNeill}},
  \bibinfo{author}{\bibfnamefont{C.}~\bibnamefont{Heikes}},
  \bibinfo{author}{\bibfnamefont{K.~F.} \bibnamefont{Mak}},
  \bibinfo{author}{\bibfnamefont{Z.}~\bibnamefont{Anderson}},
  \bibinfo{author}{\bibfnamefont{A.}~\bibnamefont{Korm{\'a}nyos}},
  \bibinfo{author}{\bibfnamefont{V.}~\bibnamefont{Z{\'o}lyomi}},
  \bibinfo{author}{\bibfnamefont{J.}~\bibnamefont{Park}}, \bibnamefont{and}
  \bibinfo{author}{\bibfnamefont{D.~C.} \bibnamefont{Ralph}},
  \bibinfo{journal}{Physical review letters} \textbf{\bibinfo{volume}{114}},
  \bibinfo{pages}{037401} (\bibinfo{year}{2015}).

\bibitem[{\citenamefont{Mitioglu et~al.}(2015)\citenamefont{Mitioglu,
  Plochocka, Granados~del Aguila, Christianen, Deligeorgis, Anghel, Kulyuk, and
  Maude}}]{Mitioglu2015}
\bibinfo{author}{\bibfnamefont{A.~A.} \bibnamefont{Mitioglu}},
  \bibinfo{author}{\bibfnamefont{P.}~\bibnamefont{Plochocka}},
  \bibinfo{author}{\bibfnamefont{A.}~\bibnamefont{Granados~del Aguila}},
  \bibinfo{author}{\bibfnamefont{P.~C.~M.} \bibnamefont{Christianen}},
  \bibinfo{author}{\bibfnamefont{G.}~\bibnamefont{Deligeorgis}},
  \bibinfo{author}{\bibfnamefont{S.}~\bibnamefont{Anghel}},
  \bibinfo{author}{\bibfnamefont{L.}~\bibnamefont{Kulyuk}}, \bibnamefont{and}
  \bibinfo{author}{\bibfnamefont{D.~K.} \bibnamefont{Maude}},
  \bibinfo{journal}{Nano Letters} \textbf{\bibinfo{volume}{15}},
  \bibinfo{pages}{4387} (\bibinfo{year}{2015}).

\bibitem[{\citenamefont{Stier et~al.}(2016)\citenamefont{Stier, McCreary,
  Jonker, Kono, and Crooker}}]{Stier2016}
\bibinfo{author}{\bibfnamefont{A.~V.} \bibnamefont{Stier}},
  \bibinfo{author}{\bibfnamefont{K.~M.} \bibnamefont{McCreary}},
  \bibinfo{author}{\bibfnamefont{B.~T.} \bibnamefont{Jonker}},
  \bibinfo{author}{\bibfnamefont{J.}~\bibnamefont{Kono}}, \bibnamefont{and}
  \bibinfo{author}{\bibfnamefont{S.~A.} \bibnamefont{Crooker}},
  \bibinfo{journal}{Nat. Commun.} \textbf{\bibinfo{volume}{7}},
  \bibinfo{pages}{10643} (\bibinfo{year}{2016}).

\bibitem[{\citenamefont{Goryca et~al.}(2019)\citenamefont{Goryca, Li, Stier,
  Taniguchi, Watanabe, Courtade, Shree, Robert, Urbaszek, Marie
  et~al.}}]{Goryca2019}
\bibinfo{author}{\bibfnamefont{M.}~\bibnamefont{Goryca}},
  \bibinfo{author}{\bibfnamefont{J.}~\bibnamefont{Li}},
  \bibinfo{author}{\bibfnamefont{A.~V.} \bibnamefont{Stier}},
  \bibinfo{author}{\bibfnamefont{T.}~\bibnamefont{Taniguchi}},
  \bibinfo{author}{\bibfnamefont{K.}~\bibnamefont{Watanabe}},
  \bibinfo{author}{\bibfnamefont{E.}~\bibnamefont{Courtade}},
  \bibinfo{author}{\bibfnamefont{S.}~\bibnamefont{Shree}},
  \bibinfo{author}{\bibfnamefont{C.}~\bibnamefont{Robert}},
  \bibinfo{author}{\bibfnamefont{B.}~\bibnamefont{Urbaszek}},
  \bibinfo{author}{\bibfnamefont{X.}~\bibnamefont{Marie}},
  \bibnamefont{et~al.}, \bibinfo{journal}{Nature Communications}
  \textbf{\bibinfo{volume}{10}}, \bibinfo{pages}{4172} (\bibinfo{year}{2019}).

\bibitem[{\citenamefont{{Zhang, Xiao-Xiao} et~al.}(2017)\citenamefont{{Zhang,
  Xiao-Xiao}, {Cao, Ting}, {Lu, Zhengguang}, {Lin, Yu-Chuan}, {Zhang, Fan},
  {Wang, Ying}, {Li, Zhiqiang}, {Hone, James C.}, {Robinson, Joshua A.},
  {Smirnov, Dmitry} et~al.}}]{Zhang2017}
\bibinfo{author}{\bibnamefont{{Zhang, Xiao-Xiao}}},
  \bibinfo{author}{\bibnamefont{{Cao, Ting}}},
  \bibinfo{author}{\bibnamefont{{Lu, Zhengguang}}},
  \bibinfo{author}{\bibnamefont{{Lin, Yu-Chuan}}},
  \bibinfo{author}{\bibnamefont{{Zhang, Fan}}},
  \bibinfo{author}{\bibnamefont{{Wang, Ying}}},
  \bibinfo{author}{\bibnamefont{{Li, Zhiqiang}}},
  \bibinfo{author}{\bibnamefont{{Hone, James C.}}},
  \bibinfo{author}{\bibnamefont{{Robinson, Joshua A.}}},
  \bibinfo{author}{\bibnamefont{{Smirnov, Dmitry}}}, \bibnamefont{et~al.},
  \bibinfo{journal}{Nature Nanotechnology} \textbf{\bibinfo{volume}{12}},
  \bibinfo{pages}{883} (\bibinfo{year}{2017}).

\bibitem[{\citenamefont{Stier et~al.}(2018)\citenamefont{Stier, Wilson,
  Velizhanin, Kono, Xu, and Crooker}}]{Stier18}
\bibinfo{author}{\bibfnamefont{A.~V.} \bibnamefont{Stier}},
  \bibinfo{author}{\bibfnamefont{N.~P.} \bibnamefont{Wilson}},
  \bibinfo{author}{\bibfnamefont{K.~A.} \bibnamefont{Velizhanin}},
  \bibinfo{author}{\bibfnamefont{J.}~\bibnamefont{Kono}},
  \bibinfo{author}{\bibfnamefont{X.}~\bibnamefont{Xu}}, \bibnamefont{and}
  \bibinfo{author}{\bibfnamefont{S.~A.} \bibnamefont{Crooker}},
  \bibinfo{journal}{Phys. Rev. Lett.} \textbf{\bibinfo{volume}{120}},
  \bibinfo{pages}{057405} (\bibinfo{year}{2018}).

\bibitem[{\citenamefont{Wang et~al.}(2020)\citenamefont{Wang, Li, Li, Lu, Miao,
  Lian, Meng, Blei, Taniguchi, Watanabe et~al.}}]{Shi-Rydberg}
\bibinfo{author}{\bibfnamefont{T.}~\bibnamefont{Wang}},
  \bibinfo{author}{\bibfnamefont{Z.}~\bibnamefont{Li}},
  \bibinfo{author}{\bibfnamefont{Y.}~\bibnamefont{Li}},
  \bibinfo{author}{\bibfnamefont{Z.}~\bibnamefont{Lu}},
  \bibinfo{author}{\bibfnamefont{S.}~\bibnamefont{Miao}},
  \bibinfo{author}{\bibfnamefont{Z.}~\bibnamefont{Lian}},
  \bibinfo{author}{\bibfnamefont{Y.}~\bibnamefont{Meng}},
  \bibinfo{author}{\bibfnamefont{M.}~\bibnamefont{Blei}},
  \bibinfo{author}{\bibfnamefont{T.}~\bibnamefont{Taniguchi}},
  \bibinfo{author}{\bibfnamefont{K.}~\bibnamefont{Watanabe}},
  \bibnamefont{et~al.}, \bibinfo{journal}{Nano Letters}
  \textbf{\bibinfo{volume}{20}}, \bibinfo{pages}{7635} (\bibinfo{year}{2020}).

\bibitem[{\citenamefont{Nagler et~al.}(2018)\citenamefont{Nagler, Ballottin,
  Mitioglu, Durnev, Taniguchi, Watanabe, Chernikov, Sch\"uller, Glazov,
  Christianen et~al.}}]{Nagler18}
\bibinfo{author}{\bibfnamefont{P.}~\bibnamefont{Nagler}},
  \bibinfo{author}{\bibfnamefont{M.~V.} \bibnamefont{Ballottin}},
  \bibinfo{author}{\bibfnamefont{A.~A.} \bibnamefont{Mitioglu}},
  \bibinfo{author}{\bibfnamefont{M.~V.} \bibnamefont{Durnev}},
  \bibinfo{author}{\bibfnamefont{T.}~\bibnamefont{Taniguchi}},
  \bibinfo{author}{\bibfnamefont{K.}~\bibnamefont{Watanabe}},
  \bibinfo{author}{\bibfnamefont{A.}~\bibnamefont{Chernikov}},
  \bibinfo{author}{\bibfnamefont{C.}~\bibnamefont{Sch\"uller}},
  \bibinfo{author}{\bibfnamefont{M.~M.} \bibnamefont{Glazov}},
  \bibinfo{author}{\bibfnamefont{P.~C.~M.} \bibnamefont{Christianen}},
  \bibnamefont{et~al.}, \bibinfo{journal}{Phys. Rev. Lett.}
  \textbf{\bibinfo{volume}{121}}, \bibinfo{pages}{057402}
  (\bibinfo{year}{2018}).

\bibitem[{\citenamefont{Barbone et~al.}(2018)\citenamefont{Barbone, Montblanch,
  Kara, Palacios-Berraquero, Cadore, De~Fazio, Pingault, Mostaani, Li, Chen
  et~al.}}]{Barbone2018}
\bibinfo{author}{\bibfnamefont{M.}~\bibnamefont{Barbone}},
  \bibinfo{author}{\bibfnamefont{A.~R.-P.} \bibnamefont{Montblanch}},
  \bibinfo{author}{\bibfnamefont{D.~M.} \bibnamefont{Kara}},
  \bibinfo{author}{\bibfnamefont{C.}~\bibnamefont{Palacios-Berraquero}},
  \bibinfo{author}{\bibfnamefont{A.~R.} \bibnamefont{Cadore}},
  \bibinfo{author}{\bibfnamefont{D.}~\bibnamefont{De~Fazio}},
  \bibinfo{author}{\bibfnamefont{B.}~\bibnamefont{Pingault}},
  \bibinfo{author}{\bibfnamefont{E.}~\bibnamefont{Mostaani}},
  \bibinfo{author}{\bibfnamefont{H.}~\bibnamefont{Li}},
  \bibinfo{author}{\bibfnamefont{B.}~\bibnamefont{Chen}}, \bibnamefont{et~al.},
  \bibinfo{journal}{Nature Communications} \textbf{\bibinfo{volume}{9}},
  \bibinfo{pages}{3721} (\bibinfo{year}{2018}).

\bibitem[{\citenamefont{Li et~al.}(2018)\citenamefont{Li, Wang, Lu, Jin, Chen,
  Meng, Lian, Taniguchi, Watanabe, Zhang et~al.}}]{Li2018}
\bibinfo{author}{\bibfnamefont{Z.}~\bibnamefont{Li}},
  \bibinfo{author}{\bibfnamefont{T.}~\bibnamefont{Wang}},
  \bibinfo{author}{\bibfnamefont{Z.}~\bibnamefont{Lu}},
  \bibinfo{author}{\bibfnamefont{C.}~\bibnamefont{Jin}},
  \bibinfo{author}{\bibfnamefont{Y.}~\bibnamefont{Chen}},
  \bibinfo{author}{\bibfnamefont{Y.}~\bibnamefont{Meng}},
  \bibinfo{author}{\bibfnamefont{Z.}~\bibnamefont{Lian}},
  \bibinfo{author}{\bibfnamefont{T.}~\bibnamefont{Taniguchi}},
  \bibinfo{author}{\bibfnamefont{K.}~\bibnamefont{Watanabe}},
  \bibinfo{author}{\bibfnamefont{S.}~\bibnamefont{Zhang}},
  \bibnamefont{et~al.}, \bibinfo{journal}{Nature Communications}
  \textbf{\bibinfo{volume}{9}}, \bibinfo{pages}{3719} (\bibinfo{year}{2018}).

\bibitem[{\citenamefont{Arora}(2021)}]{doi:10.1063/5.0042683}
\bibinfo{author}{\bibfnamefont{A.}~\bibnamefont{Arora}},
  \bibinfo{journal}{Journal of Applied Physics} \textbf{\bibinfo{volume}{129}},
  \bibinfo{pages}{120902} (\bibinfo{year}{2021}).

\bibitem[{\citenamefont{Nagler et~al.}(2017{\natexlab{b}})\citenamefont{Nagler,
  Ballottin, Mitioglu, Mooshammer, Paradiso, Strunk, Huber, Chernikov,
  Christianen, Sch\"{u}ller et~al.}}]{Nagler17b}
\bibinfo{author}{\bibfnamefont{P.}~\bibnamefont{Nagler}},
  \bibinfo{author}{\bibfnamefont{M.~V.} \bibnamefont{Ballottin}},
  \bibinfo{author}{\bibfnamefont{A.~A.} \bibnamefont{Mitioglu}},
  \bibinfo{author}{\bibfnamefont{F.}~\bibnamefont{Mooshammer}},
  \bibinfo{author}{\bibfnamefont{N.}~\bibnamefont{Paradiso}},
  \bibinfo{author}{\bibfnamefont{C.}~\bibnamefont{Strunk}},
  \bibinfo{author}{\bibfnamefont{R.}~\bibnamefont{Huber}},
  \bibinfo{author}{\bibfnamefont{A.}~\bibnamefont{Chernikov}},
  \bibinfo{author}{\bibfnamefont{P.~C.~M.} \bibnamefont{Christianen}},
  \bibinfo{author}{\bibfnamefont{C.}~\bibnamefont{Sch\"{u}ller}},
  \bibnamefont{et~al.}, \bibinfo{journal}{Nat. Commun.}
  \textbf{\bibinfo{volume}{8}}, \bibinfo{pages}{1551}
  (\bibinfo{year}{2017}{\natexlab{b}}).

\bibitem[{\citenamefont{Ciarrocchi et~al.}(2019)\citenamefont{Ciarrocchi,
  Unuchek, Avsar, Watanabe, Taniguchi, and Kis}}]{Ciarrocchi2019}
\bibinfo{author}{\bibfnamefont{A.}~\bibnamefont{Ciarrocchi}},
  \bibinfo{author}{\bibfnamefont{D.}~\bibnamefont{Unuchek}},
  \bibinfo{author}{\bibfnamefont{A.}~\bibnamefont{Avsar}},
  \bibinfo{author}{\bibfnamefont{K.}~\bibnamefont{Watanabe}},
  \bibinfo{author}{\bibfnamefont{T.}~\bibnamefont{Taniguchi}},
  \bibnamefont{and} \bibinfo{author}{\bibfnamefont{A.}~\bibnamefont{Kis}},
  \bibinfo{journal}{Nat. Photonics} \textbf{\bibinfo{volume}{13}},
  \bibinfo{pages}{131} (\bibinfo{year}{2019}).

\bibitem[{\citenamefont{Seyler et~al.}(2019)\citenamefont{Seyler, Rivera, Yu,
  Wilson, Ray, Mandrus, Yan, Yao, and Xu}}]{Seyler2019}
\bibinfo{author}{\bibfnamefont{K.~L.} \bibnamefont{Seyler}},
  \bibinfo{author}{\bibfnamefont{P.}~\bibnamefont{Rivera}},
  \bibinfo{author}{\bibfnamefont{H.}~\bibnamefont{Yu}},
  \bibinfo{author}{\bibfnamefont{N.~P.} \bibnamefont{Wilson}},
  \bibinfo{author}{\bibfnamefont{E.~L.} \bibnamefont{Ray}},
  \bibinfo{author}{\bibfnamefont{D.~G.} \bibnamefont{Mandrus}},
  \bibinfo{author}{\bibfnamefont{J.}~\bibnamefont{Yan}},
  \bibinfo{author}{\bibfnamefont{W.}~\bibnamefont{Yao}}, \bibnamefont{and}
  \bibinfo{author}{\bibfnamefont{X.}~\bibnamefont{Xu}},
  \bibinfo{journal}{Nature} \textbf{\bibinfo{volume}{567}}, \bibinfo{pages}{66}
  (\bibinfo{year}{2019}).

\bibitem[{\citenamefont{Yu et~al.}(2017)\citenamefont{Yu, Liu, Tang, Xu, and
  Yao}}]{yu2017moire}
\bibinfo{author}{\bibfnamefont{H.}~\bibnamefont{Yu}},
  \bibinfo{author}{\bibfnamefont{G.-B.} \bibnamefont{Liu}},
  \bibinfo{author}{\bibfnamefont{J.}~\bibnamefont{Tang}},
  \bibinfo{author}{\bibfnamefont{X.}~\bibnamefont{Xu}}, \bibnamefont{and}
  \bibinfo{author}{\bibfnamefont{W.}~\bibnamefont{Yao}},
  \bibinfo{journal}{Science Advances} \textbf{\bibinfo{volume}{3}},
  \bibinfo{pages}{e1701696} (\bibinfo{year}{2017}).

\bibitem[{\citenamefont{Tran et~al.}(2019)\citenamefont{Tran, Moody, Wu, Lu,
  Choi, Kim, Rai, Sanchez, Quan, Singh et~al.}}]{Tran2019}
\bibinfo{author}{\bibfnamefont{K.}~\bibnamefont{Tran}},
  \bibinfo{author}{\bibfnamefont{G.}~\bibnamefont{Moody}},
  \bibinfo{author}{\bibfnamefont{F.}~\bibnamefont{Wu}},
  \bibinfo{author}{\bibfnamefont{X.}~\bibnamefont{Lu}},
  \bibinfo{author}{\bibfnamefont{J.}~\bibnamefont{Choi}},
  \bibinfo{author}{\bibfnamefont{K.}~\bibnamefont{Kim}},
  \bibinfo{author}{\bibfnamefont{A.}~\bibnamefont{Rai}},
  \bibinfo{author}{\bibfnamefont{D.~A.} \bibnamefont{Sanchez}},
  \bibinfo{author}{\bibfnamefont{J.}~\bibnamefont{Quan}},
  \bibinfo{author}{\bibfnamefont{A.}~\bibnamefont{Singh}},
  \bibnamefont{et~al.}, \bibinfo{journal}{Nature}
  \textbf{\bibinfo{volume}{567}}, \bibinfo{pages}{71} (\bibinfo{year}{2019}).

\bibitem[{\citenamefont{Rosenberger et~al.}(2020)\citenamefont{Rosenberger,
  Chuang, Phillips, Oleshko, McCreary, Sivaram, Hellberg, and
  Jonker}}]{Rosenberger20}
\bibinfo{author}{\bibfnamefont{M.~R.} \bibnamefont{Rosenberger}},
  \bibinfo{author}{\bibfnamefont{H.-J.} \bibnamefont{Chuang}},
  \bibinfo{author}{\bibfnamefont{M.}~\bibnamefont{Phillips}},
  \bibinfo{author}{\bibfnamefont{V.~P.} \bibnamefont{Oleshko}},
  \bibinfo{author}{\bibfnamefont{K.~M.} \bibnamefont{McCreary}},
  \bibinfo{author}{\bibfnamefont{S.~V.} \bibnamefont{Sivaram}},
  \bibinfo{author}{\bibfnamefont{C.~S.} \bibnamefont{Hellberg}},
  \bibnamefont{and} \bibinfo{author}{\bibfnamefont{B.~T.}
  \bibnamefont{Jonker}}, \bibinfo{journal}{ACS Nano}
  \textbf{\bibinfo{volume}{14}}, \bibinfo{pages}{4550} (\bibinfo{year}{2020}).

\bibitem[{\citenamefont{Weston et~al.}(2020)\citenamefont{Weston, Zou,
  Enaldiev, Summerfield, Clark, Z\'{o}lyomi, Graham, Yelgel, Magorrian, Zhou
  et~al.}}]{Weston20}
\bibinfo{author}{\bibfnamefont{A.}~\bibnamefont{Weston}},
  \bibinfo{author}{\bibfnamefont{Y.}~\bibnamefont{Zou}},
  \bibinfo{author}{\bibfnamefont{V.}~\bibnamefont{Enaldiev}},
  \bibinfo{author}{\bibfnamefont{A.}~\bibnamefont{Summerfield}},
  \bibinfo{author}{\bibfnamefont{N.}~\bibnamefont{Clark}},
  \bibinfo{author}{\bibfnamefont{V.}~\bibnamefont{Z\'{o}lyomi}},
  \bibinfo{author}{\bibfnamefont{A.}~\bibnamefont{Graham}},
  \bibinfo{author}{\bibfnamefont{C.}~\bibnamefont{Yelgel}},
  \bibinfo{author}{\bibfnamefont{S.}~\bibnamefont{Magorrian}},
  \bibinfo{author}{\bibfnamefont{M.}~\bibnamefont{Zhou}}, \bibnamefont{et~al.},
  \bibinfo{journal}{Nature Nanotechnology} \textbf{\bibinfo{volume}{15}},
  \bibinfo{pages}{592} (\bibinfo{year}{2020}).

\bibitem[{\citenamefont{Brotons-Gisbert
  et~al.}(2020)\citenamefont{Brotons-Gisbert, Baek, Molina-S{\'a}nchez,
  Campbell, Scerri, White, Watanabe, Taniguchi, Bonato, and
  Gerardot}}]{brotons2020spin}
\bibinfo{author}{\bibfnamefont{M.}~\bibnamefont{Brotons-Gisbert}},
  \bibinfo{author}{\bibfnamefont{H.}~\bibnamefont{Baek}},
  \bibinfo{author}{\bibfnamefont{A.}~\bibnamefont{Molina-S{\'a}nchez}},
  \bibinfo{author}{\bibfnamefont{A.}~\bibnamefont{Campbell}},
  \bibinfo{author}{\bibfnamefont{E.}~\bibnamefont{Scerri}},
  \bibinfo{author}{\bibfnamefont{D.}~\bibnamefont{White}},
  \bibinfo{author}{\bibfnamefont{K.}~\bibnamefont{Watanabe}},
  \bibinfo{author}{\bibfnamefont{T.}~\bibnamefont{Taniguchi}},
  \bibinfo{author}{\bibfnamefont{C.}~\bibnamefont{Bonato}}, \bibnamefont{and}
  \bibinfo{author}{\bibfnamefont{B.~D.} \bibnamefont{Gerardot}},
  \bibinfo{journal}{Nat. Mater.} \textbf{\bibinfo{volume}{19}},
  \bibinfo{pages}{630} (\bibinfo{year}{2020}).

\bibitem[{\citenamefont{Shabani et~al.}(2021)\citenamefont{Shabani, Halbertal,
  Wu, Chen, Liu, Hone, Yao, Basov, Zhu, and Pasupathy}}]{shabani2021deep}
\bibinfo{author}{\bibfnamefont{S.}~\bibnamefont{Shabani}},
  \bibinfo{author}{\bibfnamefont{D.}~\bibnamefont{Halbertal}},
  \bibinfo{author}{\bibfnamefont{W.}~\bibnamefont{Wu}},
  \bibinfo{author}{\bibfnamefont{M.}~\bibnamefont{Chen}},
  \bibinfo{author}{\bibfnamefont{S.}~\bibnamefont{Liu}},
  \bibinfo{author}{\bibfnamefont{J.}~\bibnamefont{Hone}},
  \bibinfo{author}{\bibfnamefont{W.}~\bibnamefont{Yao}},
  \bibinfo{author}{\bibfnamefont{D.~N.} \bibnamefont{Basov}},
  \bibinfo{author}{\bibfnamefont{X.}~\bibnamefont{Zhu}}, \bibnamefont{and}
  \bibinfo{author}{\bibfnamefont{A.~N.} \bibnamefont{Pasupathy}},
  \bibinfo{journal}{Nat. Phys.} \textbf{\bibinfo{volume}{17}},
  \bibinfo{pages}{720} (\bibinfo{year}{2021}).

\bibitem[{\citenamefont{Zhang et~al.}(2021)\citenamefont{Zhang, Wu, Hou, Zhang,
  Chou, Watanabe, Taniguchi, Forrest, and Deng}}]{zhang2021van}
\bibinfo{author}{\bibfnamefont{L.}~\bibnamefont{Zhang}},
  \bibinfo{author}{\bibfnamefont{F.}~\bibnamefont{Wu}},
  \bibinfo{author}{\bibfnamefont{S.}~\bibnamefont{Hou}},
  \bibinfo{author}{\bibfnamefont{Z.}~\bibnamefont{Zhang}},
  \bibinfo{author}{\bibfnamefont{Y.-H.} \bibnamefont{Chou}},
  \bibinfo{author}{\bibfnamefont{K.}~\bibnamefont{Watanabe}},
  \bibinfo{author}{\bibfnamefont{T.}~\bibnamefont{Taniguchi}},
  \bibinfo{author}{\bibfnamefont{S.~R.} \bibnamefont{Forrest}},
  \bibnamefont{and} \bibinfo{author}{\bibfnamefont{H.}~\bibnamefont{Deng}},
  \bibinfo{journal}{Nature} \textbf{\bibinfo{volume}{591}}, \bibinfo{pages}{61}
  (\bibinfo{year}{2021}).

\bibitem[{\citenamefont{Delhomme et~al.}(2020)\citenamefont{Delhomme,
  Vaclavkova, Slobodeniuk, Orlita, Potemski, Basko, Watanabe, Taniguchi, Mauro,
  Barreteau et~al.}}]{Delhomme_2020}
\bibinfo{author}{\bibfnamefont{A.}~\bibnamefont{Delhomme}},
  \bibinfo{author}{\bibfnamefont{D.}~\bibnamefont{Vaclavkova}},
  \bibinfo{author}{\bibfnamefont{A.}~\bibnamefont{Slobodeniuk}},
  \bibinfo{author}{\bibfnamefont{M.}~\bibnamefont{Orlita}},
  \bibinfo{author}{\bibfnamefont{M.}~\bibnamefont{Potemski}},
  \bibinfo{author}{\bibfnamefont{D.~M.} \bibnamefont{Basko}},
  \bibinfo{author}{\bibfnamefont{K.}~\bibnamefont{Watanabe}},
  \bibinfo{author}{\bibfnamefont{T.}~\bibnamefont{Taniguchi}},
  \bibinfo{author}{\bibfnamefont{D.}~\bibnamefont{Mauro}},
  \bibinfo{author}{\bibfnamefont{C.}~\bibnamefont{Barreteau}},
  \bibnamefont{et~al.}, \bibinfo{journal}{2D Materials}
  \textbf{\bibinfo{volume}{7}}, \bibinfo{pages}{041002} (\bibinfo{year}{2020}).

\bibitem[{\citenamefont{Ribeiro-Soares
  et~al.}(2014)\citenamefont{Ribeiro-Soares, Almeida, Barros, Araujo,
  Dresselhaus, Can\ifmmode~\mbox{\c{c}}\else \c{c}\fi{}ado, and
  Jorio}}]{PhysRevB.90.115438}
\bibinfo{author}{\bibfnamefont{J.}~\bibnamefont{Ribeiro-Soares}},
  \bibinfo{author}{\bibfnamefont{R.~M.} \bibnamefont{Almeida}},
  \bibinfo{author}{\bibfnamefont{E.~B.} \bibnamefont{Barros}},
  \bibinfo{author}{\bibfnamefont{P.~T.} \bibnamefont{Araujo}},
  \bibinfo{author}{\bibfnamefont{M.~S.} \bibnamefont{Dresselhaus}},
  \bibinfo{author}{\bibfnamefont{L.~G.}
  \bibnamefont{Can\ifmmode~\mbox{\c{c}}\else \c{c}\fi{}ado}}, \bibnamefont{and}
  \bibinfo{author}{\bibfnamefont{A.}~\bibnamefont{Jorio}},
  \bibinfo{journal}{Phys. Rev. B} \textbf{\bibinfo{volume}{90}},
  \bibinfo{pages}{115438} (\bibinfo{year}{2014}).

\bibitem[{\citenamefont{Zhang and Niu}(2015)}]{PhysRevLett.115.115502}
\bibinfo{author}{\bibfnamefont{L.}~\bibnamefont{Zhang}} \bibnamefont{and}
  \bibinfo{author}{\bibfnamefont{Q.}~\bibnamefont{Niu}},
  \bibinfo{journal}{Phys. Rev. Lett.} \textbf{\bibinfo{volume}{115}},
  \bibinfo{pages}{115502} (\bibinfo{year}{2015}).

\bibitem[{\citenamefont{He et~al.}(2020)\citenamefont{He, Rivera, Van~Tuan,
  Wilson, Yang, Taniguchi, Watanabe, Yan, Mandrus, Yu et~al.}}]{he2020valley}
\bibinfo{author}{\bibfnamefont{M.}~\bibnamefont{He}},
  \bibinfo{author}{\bibfnamefont{P.}~\bibnamefont{Rivera}},
  \bibinfo{author}{\bibfnamefont{D.}~\bibnamefont{Van~Tuan}},
  \bibinfo{author}{\bibfnamefont{N.~P.} \bibnamefont{Wilson}},
  \bibinfo{author}{\bibfnamefont{M.}~\bibnamefont{Yang}},
  \bibinfo{author}{\bibfnamefont{T.}~\bibnamefont{Taniguchi}},
  \bibinfo{author}{\bibfnamefont{K.}~\bibnamefont{Watanabe}},
  \bibinfo{author}{\bibfnamefont{J.}~\bibnamefont{Yan}},
  \bibinfo{author}{\bibfnamefont{D.~G.} \bibnamefont{Mandrus}},
  \bibinfo{author}{\bibfnamefont{H.}~\bibnamefont{Yu}}, \bibnamefont{et~al.},
  \bibinfo{journal}{Nat. Commun.} \textbf{\bibinfo{volume}{11}},
  \bibinfo{pages}{1} (\bibinfo{year}{2020}).

\bibitem[{\citenamefont{Kaasbjerg et~al.}(2012)\citenamefont{Kaasbjerg,
  Thygesen, and Jacobsen}}]{Kaasbjerg2012}
\bibinfo{author}{\bibfnamefont{K.}~\bibnamefont{Kaasbjerg}},
  \bibinfo{author}{\bibfnamefont{K.~S.} \bibnamefont{Thygesen}},
  \bibnamefont{and} \bibinfo{author}{\bibfnamefont{K.~W.}
  \bibnamefont{Jacobsen}}, \bibinfo{journal}{Phys. Rev. B}
  \textbf{\bibinfo{volume}{85}}, \bibinfo{pages}{115317}
  (\bibinfo{year}{2012}).

\bibitem[{\citenamefont{Song and Dery}(2013)}]{Song2013}
\bibinfo{author}{\bibfnamefont{Y.}~\bibnamefont{Song}} \bibnamefont{and}
  \bibinfo{author}{\bibfnamefont{H.}~\bibnamefont{Dery}},
  \bibinfo{journal}{Phys. Rev. Lett.} \textbf{\bibinfo{volume}{111}},
  \bibinfo{pages}{026601} (\bibinfo{year}{2013}).

\bibitem[{\citenamefont{Li et~al.}(2013)\citenamefont{Li, Mullen, Jin,
  Borysenko, Buongiorno~Nardelli, and Kim}}]{PhysRevB.87.115418}
\bibinfo{author}{\bibfnamefont{X.}~\bibnamefont{Li}},
  \bibinfo{author}{\bibfnamefont{J.~T.} \bibnamefont{Mullen}},
  \bibinfo{author}{\bibfnamefont{Z.}~\bibnamefont{Jin}},
  \bibinfo{author}{\bibfnamefont{K.~M.} \bibnamefont{Borysenko}},
  \bibinfo{author}{\bibfnamefont{M.}~\bibnamefont{Buongiorno~Nardelli}},
  \bibnamefont{and} \bibinfo{author}{\bibfnamefont{K.~W.} \bibnamefont{Kim}},
  \bibinfo{journal}{Phys. Rev. B} \textbf{\bibinfo{volume}{87}},
  \bibinfo{pages}{115418} (\bibinfo{year}{2013}).

\bibitem[{\citenamefont{Jin et~al.}(2014)\citenamefont{Jin, Li, Mullen, and
  Kim}}]{Jin2014}
\bibinfo{author}{\bibfnamefont{Z.}~\bibnamefont{Jin}},
  \bibinfo{author}{\bibfnamefont{X.}~\bibnamefont{Li}},
  \bibinfo{author}{\bibfnamefont{J.~T.} \bibnamefont{Mullen}},
  \bibnamefont{and} \bibinfo{author}{\bibfnamefont{K.~W.} \bibnamefont{Kim}},
  \bibinfo{journal}{Phys. Rev. B} \textbf{\bibinfo{volume}{90}},
  \bibinfo{pages}{045422} (\bibinfo{year}{2014}).

\bibitem[{\citenamefont{Christiansen et~al.}(2017)\citenamefont{Christiansen,
  Selig, Bergh\"auser, Schmidt, Niehues, Schneider, Arora, de~Vasconcellos,
  Bratschitsch, Malic et~al.}}]{Christiansen2017}
\bibinfo{author}{\bibfnamefont{D.}~\bibnamefont{Christiansen}},
  \bibinfo{author}{\bibfnamefont{M.}~\bibnamefont{Selig}},
  \bibinfo{author}{\bibfnamefont{G.}~\bibnamefont{Bergh\"auser}},
  \bibinfo{author}{\bibfnamefont{R.}~\bibnamefont{Schmidt}},
  \bibinfo{author}{\bibfnamefont{I.}~\bibnamefont{Niehues}},
  \bibinfo{author}{\bibfnamefont{R.}~\bibnamefont{Schneider}},
  \bibinfo{author}{\bibfnamefont{A.}~\bibnamefont{Arora}},
  \bibinfo{author}{\bibfnamefont{S.~M.} \bibnamefont{de~Vasconcellos}},
  \bibinfo{author}{\bibfnamefont{R.}~\bibnamefont{Bratschitsch}},
  \bibinfo{author}{\bibfnamefont{E.}~\bibnamefont{Malic}},
  \bibnamefont{et~al.}, \bibinfo{journal}{Phys. Rev. Lett.}
  \textbf{\bibinfo{volume}{119}}, \bibinfo{pages}{187402}
  (\bibinfo{year}{2017}).

\bibitem[{\citenamefont{Li and Wang}(2018)}]{doi:10.1063/1.5030678}
\bibinfo{author}{\bibfnamefont{P.-F.} \bibnamefont{Li}} \bibnamefont{and}
  \bibinfo{author}{\bibfnamefont{Z.-W.} \bibnamefont{Wang}},
  \bibinfo{journal}{J. Appl. Phys.} \textbf{\bibinfo{volume}{123}},
  \bibinfo{pages}{204308} (\bibinfo{year}{2018}).

\bibitem[{\citenamefont{Chen et~al.}(2018)\citenamefont{Chen, Wang, and
  Peeters}}]{doi:10.1063/1.5025907}
\bibinfo{author}{\bibfnamefont{Q.}~\bibnamefont{Chen}},
  \bibinfo{author}{\bibfnamefont{W.}~\bibnamefont{Wang}}, \bibnamefont{and}
  \bibinfo{author}{\bibfnamefont{F.~M.} \bibnamefont{Peeters}},
  \bibinfo{journal}{J. Appl. Phys.} \textbf{\bibinfo{volume}{123}},
  \bibinfo{pages}{214303} (\bibinfo{year}{2018}).

\bibitem[{\citenamefont{Glazov et~al.}(2019)\citenamefont{Glazov, Semina,
  Robert, Urbaszek, Amand, and Marie}}]{PhysRevB.100.041301}
\bibinfo{author}{\bibfnamefont{M.~M.} \bibnamefont{Glazov}},
  \bibinfo{author}{\bibfnamefont{M.~A.} \bibnamefont{Semina}},
  \bibinfo{author}{\bibfnamefont{C.}~\bibnamefont{Robert}},
  \bibinfo{author}{\bibfnamefont{B.}~\bibnamefont{Urbaszek}},
  \bibinfo{author}{\bibfnamefont{T.}~\bibnamefont{Amand}}, \bibnamefont{and}
  \bibinfo{author}{\bibfnamefont{X.}~\bibnamefont{Marie}},
  \bibinfo{journal}{Phys. Rev. B} \textbf{\bibinfo{volume}{100}},
  \bibinfo{pages}{041301} (\bibinfo{year}{2019}).

\bibitem[{\citenamefont{Chow et~al.}(2017)\citenamefont{Chow, Yu, Jones,
  Schaibley, Koehler, Mandrus, Merlin, Yao, and Xu}}]{chow2017phonon}
\bibinfo{author}{\bibfnamefont{C.~M.} \bibnamefont{Chow}},
  \bibinfo{author}{\bibfnamefont{H.}~\bibnamefont{Yu}},
  \bibinfo{author}{\bibfnamefont{A.~M.} \bibnamefont{Jones}},
  \bibinfo{author}{\bibfnamefont{J.~R.} \bibnamefont{Schaibley}},
  \bibinfo{author}{\bibfnamefont{M.}~\bibnamefont{Koehler}},
  \bibinfo{author}{\bibfnamefont{D.~G.} \bibnamefont{Mandrus}},
  \bibinfo{author}{\bibfnamefont{R.}~\bibnamefont{Merlin}},
  \bibinfo{author}{\bibfnamefont{W.}~\bibnamefont{Yao}}, \bibnamefont{and}
  \bibinfo{author}{\bibfnamefont{X.}~\bibnamefont{Xu}}, \bibinfo{journal}{npj
  2D Mater. Appl.} \textbf{\bibinfo{volume}{1}}, \bibinfo{pages}{1}
  (\bibinfo{year}{2017}).

\bibitem[{\citenamefont{Shree et~al.}(2018)\citenamefont{Shree, Semina, Robert,
  Han, Amand, Balocchi, Manca, Courtade, Marie, Taniguchi
  et~al.}}]{PhysRevB.98.035302}
\bibinfo{author}{\bibfnamefont{S.}~\bibnamefont{Shree}},
  \bibinfo{author}{\bibfnamefont{M.}~\bibnamefont{Semina}},
  \bibinfo{author}{\bibfnamefont{C.}~\bibnamefont{Robert}},
  \bibinfo{author}{\bibfnamefont{B.}~\bibnamefont{Han}},
  \bibinfo{author}{\bibfnamefont{T.}~\bibnamefont{Amand}},
  \bibinfo{author}{\bibfnamefont{A.}~\bibnamefont{Balocchi}},
  \bibinfo{author}{\bibfnamefont{M.}~\bibnamefont{Manca}},
  \bibinfo{author}{\bibfnamefont{E.}~\bibnamefont{Courtade}},
  \bibinfo{author}{\bibfnamefont{X.}~\bibnamefont{Marie}},
  \bibinfo{author}{\bibfnamefont{T.}~\bibnamefont{Taniguchi}},
  \bibnamefont{et~al.}, \bibinfo{journal}{Phys. Rev. B}
  \textbf{\bibinfo{volume}{98}}, \bibinfo{pages}{035302}
  (\bibinfo{year}{2018}).

\bibitem[{\citenamefont{Brem et~al.}(2018)\citenamefont{Brem, Selig,
  Bergh\"auser, and Malic}}]{Brem2018}
\bibinfo{author}{\bibfnamefont{S.}~\bibnamefont{Brem}},
  \bibinfo{author}{\bibfnamefont{M.}~\bibnamefont{Selig}},
  \bibinfo{author}{\bibfnamefont{G.}~\bibnamefont{Bergh\"auser}},
  \bibnamefont{and} \bibinfo{author}{\bibfnamefont{E.}~\bibnamefont{Malic}},
  \bibinfo{journal}{Scientific Reports} \textbf{\bibinfo{volume}{8}},
  \bibinfo{pages}{8238} (\bibinfo{year}{2018}).

\bibitem[{\citenamefont{Paradisanos et~al.}(2021)\citenamefont{Paradisanos,
  Wang, Alexeev, Cadore, Marie, Ferrari, Glazov, and
  Urbaszek}}]{paradisanos2021efficient}
\bibinfo{author}{\bibfnamefont{I.}~\bibnamefont{Paradisanos}},
  \bibinfo{author}{\bibfnamefont{G.}~\bibnamefont{Wang}},
  \bibinfo{author}{\bibfnamefont{E.~M.} \bibnamefont{Alexeev}},
  \bibinfo{author}{\bibfnamefont{A.~R.} \bibnamefont{Cadore}},
  \bibinfo{author}{\bibfnamefont{X.}~\bibnamefont{Marie}},
  \bibinfo{author}{\bibfnamefont{A.~C.} \bibnamefont{Ferrari}},
  \bibinfo{author}{\bibfnamefont{M.~M.} \bibnamefont{Glazov}},
  \bibnamefont{and} \bibinfo{author}{\bibfnamefont{B.}~\bibnamefont{Urbaszek}},
  \bibinfo{journal}{Nat. Commun.} \textbf{\bibinfo{volume}{12}},
  \bibinfo{pages}{1} (\bibinfo{year}{2021}).

\bibitem[{\citenamefont{Gurevich and Firsov}(1961)}]{gurevich1961theory}
\bibinfo{author}{\bibfnamefont{V.~L.} \bibnamefont{Gurevich}} \bibnamefont{and}
  \bibinfo{author}{\bibfnamefont{Y.~A.} \bibnamefont{Firsov}},
  \bibinfo{journal}{Sov. Phys. JETP} \textbf{\bibinfo{volume}{13}},
  \bibinfo{pages}{137} (\bibinfo{year}{1961}).

\bibitem[{\citenamefont{Pavlov et~al.}(1965)\citenamefont{Pavlov, Parfen'ev,
  Firsov, and Shalyt}}]{firsov1965effect}
\bibinfo{author}{\bibfnamefont{S.}~\bibnamefont{Pavlov}},
  \bibinfo{author}{\bibfnamefont{R.}~\bibnamefont{Parfen'ev}},
  \bibinfo{author}{\bibfnamefont{Y.~A.} \bibnamefont{Firsov}},
  \bibnamefont{and} \bibinfo{author}{\bibfnamefont{S.}~\bibnamefont{Shalyt}},
  \bibinfo{journal}{Sov. Phys. JETP} \textbf{\bibinfo{volume}{21}},
  \bibinfo{pages}{1049} (\bibinfo{year}{1965}).

\bibitem[{\citenamefont{Aksel'rod and Tsidil'kovskii}(1966)}]{Akselrod1965}
\bibinfo{author}{\bibfnamefont{M.~M.} \bibnamefont{Aksel'rod}}
  \bibnamefont{and} \bibinfo{author}{\bibfnamefont{I.~M.}
  \bibnamefont{Tsidil'kovskii}}, \bibinfo{journal}{JETP Lett.}
  \textbf{\bibinfo{volume}{4}}, \bibinfo{pages}{205} (\bibinfo{year}{1966}).

\bibitem[{\citenamefont{Firsov et~al.}(1991)\citenamefont{Firsov, Gurevich,
  Parfeniev, and Tsidil'kovskii}}]{FIRSOV19911181}
\bibinfo{author}{\bibfnamefont{Y.}~\bibnamefont{Firsov}},
  \bibinfo{author}{\bibfnamefont{V.}~\bibnamefont{Gurevich}},
  \bibinfo{author}{\bibfnamefont{R.}~\bibnamefont{Parfeniev}},
  \bibnamefont{and}
  \bibinfo{author}{\bibfnamefont{I.}~\bibnamefont{Tsidil'kovskii}}, in
  \emph{\bibinfo{booktitle}{Landau Level Spectroscopy}}, edited by
  \bibinfo{editor}{\bibfnamefont{G.}~\bibnamefont{Landwehr}} \bibnamefont{and}
  \bibinfo{editor}{\bibfnamefont{E.~I.} \bibnamefont{Rashba}}
  (\bibinfo{publisher}{Elsevier}, \bibinfo{year}{1991}),
  vol.~\bibinfo{volume}{27} of \emph{\bibinfo{series}{{Modern Problems in
  Condensed Matter Sciences}}}, p. \bibinfo{pages}{1181}.

\bibitem[{\citenamefont{Basko et~al.}(2016)\citenamefont{Basko, Leszczynski,
  Faugeras, Binder, Nicolet, Kossacki, Orlita, and Potemski}}]{Basko_2016}
\bibinfo{author}{\bibfnamefont{D.~M.} \bibnamefont{Basko}},
  \bibinfo{author}{\bibfnamefont{P.}~\bibnamefont{Leszczynski}},
  \bibinfo{author}{\bibfnamefont{C.}~\bibnamefont{Faugeras}},
  \bibinfo{author}{\bibfnamefont{J.}~\bibnamefont{Binder}},
  \bibinfo{author}{\bibfnamefont{A.~A.~L.} \bibnamefont{Nicolet}},
  \bibinfo{author}{\bibfnamefont{P.}~\bibnamefont{Kossacki}},
  \bibinfo{author}{\bibfnamefont{M.}~\bibnamefont{Orlita}}, \bibnamefont{and}
  \bibinfo{author}{\bibfnamefont{M.}~\bibnamefont{Potemski}},
  \bibinfo{journal}{2D Materials} \textbf{\bibinfo{volume}{3}},
  \bibinfo{pages}{015004} (\bibinfo{year}{2016}).

\bibitem[{\citenamefont{Langerak et~al.}(1988)\citenamefont{Langerak,
  Singleton, van~der Wel, Perenboom, Barnes, Nicholas, Hopkins, and
  Foxon}}]{PhysRevB.38.13133}
\bibinfo{author}{\bibfnamefont{C.~J. G.~M.} \bibnamefont{Langerak}},
  \bibinfo{author}{\bibfnamefont{J.}~\bibnamefont{Singleton}},
  \bibinfo{author}{\bibfnamefont{P.~J.} \bibnamefont{van~der Wel}},
  \bibinfo{author}{\bibfnamefont{J.~A. A.~J.} \bibnamefont{Perenboom}},
  \bibinfo{author}{\bibfnamefont{D.~J.} \bibnamefont{Barnes}},
  \bibinfo{author}{\bibfnamefont{R.~J.} \bibnamefont{Nicholas}},
  \bibinfo{author}{\bibfnamefont{M.~A.} \bibnamefont{Hopkins}},
  \bibnamefont{and} \bibinfo{author}{\bibfnamefont{C.~T.~B.}
  \bibnamefont{Foxon}}, \bibinfo{journal}{Phys. Rev. B}
  \textbf{\bibinfo{volume}{38}}, \bibinfo{pages}{13133} (\bibinfo{year}{1988}).

\bibitem[{\citenamefont{Barnes et~al.}(1991)\citenamefont{Barnes, Nicholas,
  Peeters, Wu, Devreese, Singleton, Langerak, Harris, and
  Foxon}}]{PhysRevLett.66.794}
\bibinfo{author}{\bibfnamefont{D.~J.} \bibnamefont{Barnes}},
  \bibinfo{author}{\bibfnamefont{R.~J.} \bibnamefont{Nicholas}},
  \bibinfo{author}{\bibfnamefont{F.~M.} \bibnamefont{Peeters}},
  \bibinfo{author}{\bibfnamefont{X.-G.} \bibnamefont{Wu}},
  \bibinfo{author}{\bibfnamefont{J.~T.} \bibnamefont{Devreese}},
  \bibinfo{author}{\bibfnamefont{J.}~\bibnamefont{Singleton}},
  \bibinfo{author}{\bibfnamefont{C.~J. G.~M.} \bibnamefont{Langerak}},
  \bibinfo{author}{\bibfnamefont{J.~J.} \bibnamefont{Harris}},
  \bibnamefont{and} \bibinfo{author}{\bibfnamefont{C.~T.} \bibnamefont{Foxon}},
  \bibinfo{journal}{Phys. Rev. Lett.} \textbf{\bibinfo{volume}{66}},
  \bibinfo{pages}{794} (\bibinfo{year}{1991}).

\bibitem[{\citenamefont{Vaughan et~al.}(1996)\citenamefont{Vaughan, Nicholas,
  Langerak, Murdin, Pidgeon, Mason, and Walker}}]{PhysRevB.53.16481}
\bibinfo{author}{\bibfnamefont{T.~A.} \bibnamefont{Vaughan}},
  \bibinfo{author}{\bibfnamefont{R.~J.} \bibnamefont{Nicholas}},
  \bibinfo{author}{\bibfnamefont{C.~J. G.~M.} \bibnamefont{Langerak}},
  \bibinfo{author}{\bibfnamefont{B.~N.} \bibnamefont{Murdin}},
  \bibinfo{author}{\bibfnamefont{C.~R.} \bibnamefont{Pidgeon}},
  \bibinfo{author}{\bibfnamefont{N.~J.} \bibnamefont{Mason}}, \bibnamefont{and}
  \bibinfo{author}{\bibfnamefont{P.~J.} \bibnamefont{Walker}},
  \bibinfo{journal}{Phys. Rev. B} \textbf{\bibinfo{volume}{53}},
  \bibinfo{pages}{16481} (\bibinfo{year}{1996}).

\bibitem[{\citenamefont{Holler et~al.}(2022)\citenamefont{Holler, Selig, Kempf,
  Zipfel, Nagler, Katzer, Katsch, Ballottin, Mitioglu, Chernikov
  et~al.}}]{Holler22}
\bibinfo{author}{\bibfnamefont{J.}~\bibnamefont{Holler}},
  \bibinfo{author}{\bibfnamefont{M.}~\bibnamefont{Selig}},
  \bibinfo{author}{\bibfnamefont{M.}~\bibnamefont{Kempf}},
  \bibinfo{author}{\bibfnamefont{J.}~\bibnamefont{Zipfel}},
  \bibinfo{author}{\bibfnamefont{P.}~\bibnamefont{Nagler}},
  \bibinfo{author}{\bibfnamefont{M.}~\bibnamefont{Katzer}},
  \bibinfo{author}{\bibfnamefont{F.}~\bibnamefont{Katsch}},
  \bibinfo{author}{\bibfnamefont{M.~V.} \bibnamefont{Ballottin}},
  \bibinfo{author}{\bibfnamefont{A.~A.} \bibnamefont{Mitioglu}},
  \bibinfo{author}{\bibfnamefont{A.}~\bibnamefont{Chernikov}},
  \bibnamefont{et~al.}, \bibinfo{journal}{Phys. Rev. B}
  \textbf{\bibinfo{volume}{105}}, \bibinfo{pages}{085303}
  (\bibinfo{year}{2022}).

\bibitem[{\citenamefont{Zhang et~al.}(2019)\citenamefont{Zhang, Gogna, Burg,
  Horng, Paik, Chou, Kim, Tutuc, and Deng}}]{PhysRevB.100.041402}
\bibinfo{author}{\bibfnamefont{L.}~\bibnamefont{Zhang}},
  \bibinfo{author}{\bibfnamefont{R.}~\bibnamefont{Gogna}},
  \bibinfo{author}{\bibfnamefont{G.~W.} \bibnamefont{Burg}},
  \bibinfo{author}{\bibfnamefont{J.}~\bibnamefont{Horng}},
  \bibinfo{author}{\bibfnamefont{E.}~\bibnamefont{Paik}},
  \bibinfo{author}{\bibfnamefont{Y.-H.} \bibnamefont{Chou}},
  \bibinfo{author}{\bibfnamefont{K.}~\bibnamefont{Kim}},
  \bibinfo{author}{\bibfnamefont{E.}~\bibnamefont{Tutuc}}, \bibnamefont{and}
  \bibinfo{author}{\bibfnamefont{H.}~\bibnamefont{Deng}},
  \bibinfo{journal}{Phys. Rev. B} \textbf{\bibinfo{volume}{100}},
  \bibinfo{pages}{041402} (\bibinfo{year}{2019}).

\bibitem[{\citenamefont{Joe et~al.}(2021)\citenamefont{Joe, Jauregui,
  Pistunova, Mier~Valdivia, Lu, Wild, Scuri, De~Greve, Gelly, Zhou
  et~al.}}]{electrically-controlled}
\bibinfo{author}{\bibfnamefont{A.~Y.} \bibnamefont{Joe}},
  \bibinfo{author}{\bibfnamefont{L.~A.} \bibnamefont{Jauregui}},
  \bibinfo{author}{\bibfnamefont{K.}~\bibnamefont{Pistunova}},
  \bibinfo{author}{\bibfnamefont{A.~M.} \bibnamefont{Mier~Valdivia}},
  \bibinfo{author}{\bibfnamefont{Z.}~\bibnamefont{Lu}},
  \bibinfo{author}{\bibfnamefont{D.~S.} \bibnamefont{Wild}},
  \bibinfo{author}{\bibfnamefont{G.}~\bibnamefont{Scuri}},
  \bibinfo{author}{\bibfnamefont{K.}~\bibnamefont{De~Greve}},
  \bibinfo{author}{\bibfnamefont{R.~J.} \bibnamefont{Gelly}},
  \bibinfo{author}{\bibfnamefont{Y.}~\bibnamefont{Zhou}}, \bibnamefont{et~al.},
  \bibinfo{journal}{Phys. Rev. B} \textbf{\bibinfo{volume}{103}},
  \bibinfo{pages}{L161411} (\bibinfo{year}{2021}).

\bibitem[{\citenamefont{Wo{\'z}niak et~al.}(2020)\citenamefont{Wo{\'z}niak,
  Faria~Junior, Seifert, Chaves, and Kunstmann}}]{wozniak2020exciton}
\bibinfo{author}{\bibfnamefont{T.}~\bibnamefont{Wo{\'z}niak}},
  \bibinfo{author}{\bibfnamefont{P.~E.} \bibnamefont{Faria~Junior}},
  \bibinfo{author}{\bibfnamefont{G.}~\bibnamefont{Seifert}},
  \bibinfo{author}{\bibfnamefont{A.}~\bibnamefont{Chaves}}, \bibnamefont{and}
  \bibinfo{author}{\bibfnamefont{J.}~\bibnamefont{Kunstmann}},
  \bibinfo{journal}{Physical Review B} \textbf{\bibinfo{volume}{101}},
  \bibinfo{pages}{235408} (\bibinfo{year}{2020}).

\bibitem[{\citenamefont{Xuan and Quek}(2020)}]{PhysRevResearch.2.033256}
\bibinfo{author}{\bibfnamefont{F.}~\bibnamefont{Xuan}} \bibnamefont{and}
  \bibinfo{author}{\bibfnamefont{S.~Y.} \bibnamefont{Quek}},
  \bibinfo{journal}{Phys. Rev. Research} \textbf{\bibinfo{volume}{2}},
  \bibinfo{pages}{033256} (\bibinfo{year}{2020}).

\bibitem[{\citenamefont{Deilmann et~al.}(2020)\citenamefont{Deilmann, Kr\"uger,
  and Rohlfing}}]{Deilmann2020}
\bibinfo{author}{\bibfnamefont{T.}~\bibnamefont{Deilmann}},
  \bibinfo{author}{\bibfnamefont{P.}~\bibnamefont{Kr\"uger}}, \bibnamefont{and}
  \bibinfo{author}{\bibfnamefont{M.}~\bibnamefont{Rohlfing}},
  \bibinfo{journal}{Phys. Rev. Lett.} \textbf{\bibinfo{volume}{124}},
  \bibinfo{pages}{226402} (\bibinfo{year}{2020}).

\bibitem[{\citenamefont{F{\"o}rste et~al.}(2020)\citenamefont{F{\"o}rste,
  Tepliakov, Kruchinin, Lindlau, Funk, F{\"o}rg, Watanabe, Taniguchi,
  Baimuratov, and H{\"o}gele}}]{Forste2020a}
\bibinfo{author}{\bibfnamefont{J.}~\bibnamefont{F{\"o}rste}},
  \bibinfo{author}{\bibfnamefont{N.~V.} \bibnamefont{Tepliakov}},
  \bibinfo{author}{\bibfnamefont{S.~Y.} \bibnamefont{Kruchinin}},
  \bibinfo{author}{\bibfnamefont{J.}~\bibnamefont{Lindlau}},
  \bibinfo{author}{\bibfnamefont{V.}~\bibnamefont{Funk}},
  \bibinfo{author}{\bibfnamefont{M.}~\bibnamefont{F{\"o}rg}},
  \bibinfo{author}{\bibfnamefont{K.}~\bibnamefont{Watanabe}},
  \bibinfo{author}{\bibfnamefont{T.}~\bibnamefont{Taniguchi}},
  \bibinfo{author}{\bibfnamefont{A.~S.} \bibnamefont{Baimuratov}},
  \bibnamefont{and}
  \bibinfo{author}{\bibfnamefont{A.}~\bibnamefont{H{\"o}gele}},
  \bibinfo{journal}{Nature Communications} \textbf{\bibinfo{volume}{11}},
  \bibinfo{pages}{4539} (\bibinfo{year}{2020}).

\bibitem[{\citenamefont{Horzum et~al.}(2013)\citenamefont{Horzum, Sahin,
  Cahangirov, Cudazzo, Rubio, Serin, and Peeters}}]{Horzum2013}
\bibinfo{author}{\bibfnamefont{S.}~\bibnamefont{Horzum}},
  \bibinfo{author}{\bibfnamefont{H.}~\bibnamefont{Sahin}},
  \bibinfo{author}{\bibfnamefont{S.}~\bibnamefont{Cahangirov}},
  \bibinfo{author}{\bibfnamefont{P.}~\bibnamefont{Cudazzo}},
  \bibinfo{author}{\bibfnamefont{a.}~\bibnamefont{Rubio}},
  \bibinfo{author}{\bibfnamefont{T.}~\bibnamefont{Serin}}, \bibnamefont{and}
  \bibinfo{author}{\bibfnamefont{F.~M.} \bibnamefont{Peeters}},
  \bibinfo{journal}{Phys. Rev. B} \textbf{\bibinfo{volume}{87}},
  \bibinfo{pages}{1} (\bibinfo{year}{2013}), ISSN \bibinfo{issn}{10980121},
  \eprint{1302.6635}.

\bibitem[{\citenamefont{Huang et~al.}(2013)\citenamefont{Huang, Da, and
  Liang}}]{doi:10.1063/1.4794363}
\bibinfo{author}{\bibfnamefont{W.}~\bibnamefont{Huang}},
  \bibinfo{author}{\bibfnamefont{H.}~\bibnamefont{Da}}, \bibnamefont{and}
  \bibinfo{author}{\bibfnamefont{G.}~\bibnamefont{Liang}},
  \bibinfo{journal}{Journal of Applied Physics} \textbf{\bibinfo{volume}{113}},
  \bibinfo{pages}{104304} (\bibinfo{year}{2013}).

\bibitem[{\citenamefont{Peng et~al.}(2016)\citenamefont{Peng, Zhang, Shao, Xu,
  Zhang, and Zhu}}]{C5RA19747C}
\bibinfo{author}{\bibfnamefont{B.}~\bibnamefont{Peng}},
  \bibinfo{author}{\bibfnamefont{H.}~\bibnamefont{Zhang}},
  \bibinfo{author}{\bibfnamefont{H.}~\bibnamefont{Shao}},
  \bibinfo{author}{\bibfnamefont{Y.}~\bibnamefont{Xu}},
  \bibinfo{author}{\bibfnamefont{X.}~\bibnamefont{Zhang}}, \bibnamefont{and}
  \bibinfo{author}{\bibfnamefont{H.}~\bibnamefont{Zhu}}, \bibinfo{journal}{RSC
  Adv.} \textbf{\bibinfo{volume}{6}}, \bibinfo{pages}{5767}
  (\bibinfo{year}{2016}).

\bibitem[{\citenamefont{Lin et~al.}(2021{\natexlab{a}})\citenamefont{Lin, Ong,
  Bange, Faria~J., Peng, Ziegler, Zipfel, B{\"a}uml, Paradiso, Watanabe
  et~al.}}]{lin2021narrow}
\bibinfo{author}{\bibfnamefont{K.-Q.} \bibnamefont{Lin}},
  \bibinfo{author}{\bibfnamefont{C.~S.} \bibnamefont{Ong}},
  \bibinfo{author}{\bibfnamefont{S.}~\bibnamefont{Bange}},
  \bibinfo{author}{\bibfnamefont{P.~E.} \bibnamefont{Faria~J.}},
  \bibinfo{author}{\bibfnamefont{B.}~\bibnamefont{Peng}},
  \bibinfo{author}{\bibfnamefont{J.~D.} \bibnamefont{Ziegler}},
  \bibinfo{author}{\bibfnamefont{J.}~\bibnamefont{Zipfel}},
  \bibinfo{author}{\bibfnamefont{C.}~\bibnamefont{B{\"a}uml}},
  \bibinfo{author}{\bibfnamefont{N.}~\bibnamefont{Paradiso}},
  \bibinfo{author}{\bibfnamefont{K.}~\bibnamefont{Watanabe}},
  \bibnamefont{et~al.}, \bibinfo{journal}{Nat. Commun.}
  \textbf{\bibinfo{volume}{12}}, \bibinfo{pages}{1}
  (\bibinfo{year}{2021}{\natexlab{a}}).

\bibitem[{\citenamefont{Lin et~al.}(2021{\natexlab{b}})\citenamefont{Lin,
  Holler, Bauer, Parzefall, Scheuck, Peng, Korn, Bange, Lupton, and
  Schüller}}]{Lin21}
\bibinfo{author}{\bibfnamefont{K.-Q.} \bibnamefont{Lin}},
  \bibinfo{author}{\bibfnamefont{J.}~\bibnamefont{Holler}},
  \bibinfo{author}{\bibfnamefont{J.~M.} \bibnamefont{Bauer}},
  \bibinfo{author}{\bibfnamefont{P.}~\bibnamefont{Parzefall}},
  \bibinfo{author}{\bibfnamefont{M.}~\bibnamefont{Scheuck}},
  \bibinfo{author}{\bibfnamefont{B.}~\bibnamefont{Peng}},
  \bibinfo{author}{\bibfnamefont{T.}~\bibnamefont{Korn}},
  \bibinfo{author}{\bibfnamefont{S.}~\bibnamefont{Bange}},
  \bibinfo{author}{\bibfnamefont{J.~M.} \bibnamefont{Lupton}},
  \bibnamefont{and}
  \bibinfo{author}{\bibfnamefont{C.}~\bibnamefont{Schüller}},
  \bibinfo{journal}{Advanced Materials} \textbf{\bibinfo{volume}{33}},
  \bibinfo{pages}{2008333} (\bibinfo{year}{2021}{\natexlab{b}}).

\bibitem[{\citenamefont{Mahrouche et~al.}(2022)\citenamefont{Mahrouche,
  Rezouali, Mahtout, Zaabar, and
  Molina-Sánchez}}]{https://doi.org/10.1002/pssb.202100321}
\bibinfo{author}{\bibfnamefont{F.}~\bibnamefont{Mahrouche}},
  \bibinfo{author}{\bibfnamefont{K.}~\bibnamefont{Rezouali}},
  \bibinfo{author}{\bibfnamefont{S.}~\bibnamefont{Mahtout}},
  \bibinfo{author}{\bibfnamefont{F.}~\bibnamefont{Zaabar}}, \bibnamefont{and}
  \bibinfo{author}{\bibfnamefont{A.}~\bibnamefont{Molina-Sánchez}},
  \bibinfo{journal}{Physica Status Solidi (B)} \textbf{\bibinfo{volume}{259}},
  \bibinfo{pages}{2100321} (\bibinfo{year}{2022}).

\bibitem[{\citenamefont{Parzefall et~al.}(2021)\citenamefont{Parzefall, Holler,
  Scheuck, Beer, Lin, Peng, Monserrat, Nagler, Kempf, Korn
  et~al.}}]{Parzefall_2021}
\bibinfo{author}{\bibfnamefont{P.}~\bibnamefont{Parzefall}},
  \bibinfo{author}{\bibfnamefont{J.}~\bibnamefont{Holler}},
  \bibinfo{author}{\bibfnamefont{M.}~\bibnamefont{Scheuck}},
  \bibinfo{author}{\bibfnamefont{A.}~\bibnamefont{Beer}},
  \bibinfo{author}{\bibfnamefont{K.-Q.} \bibnamefont{Lin}},
  \bibinfo{author}{\bibfnamefont{B.}~\bibnamefont{Peng}},
  \bibinfo{author}{\bibfnamefont{B.}~\bibnamefont{Monserrat}},
  \bibinfo{author}{\bibfnamefont{P.}~\bibnamefont{Nagler}},
  \bibinfo{author}{\bibfnamefont{M.}~\bibnamefont{Kempf}},
  \bibinfo{author}{\bibfnamefont{T.}~\bibnamefont{Korn}}, \bibnamefont{et~al.},
  \bibinfo{journal}{2D Materials} \textbf{\bibinfo{volume}{8}},
  \bibinfo{pages}{035030} (\bibinfo{year}{2021}).

\bibitem[{SI()}]{SI}
\emph{\bibinfo{title}{\normalfont{See Supplementary material for the details on
  the modelling of the PL polarization degree, calculation of the resonant
  intervalley scattering time, and supplementary discussion of it.}}}

\bibitem[{\citenamefont{Pearce and Burkard}(2017)}]{Pearce2017}
\bibinfo{author}{\bibfnamefont{A.~J.} \bibnamefont{Pearce}} \bibnamefont{and}
  \bibinfo{author}{\bibfnamefont{G.}~\bibnamefont{Burkard}},
  \bibinfo{journal}{2D Mater.} \textbf{\bibinfo{volume}{4}},
  \bibinfo{pages}{025114} (\bibinfo{year}{2017}).

\bibitem[{\citenamefont{Bir and Pikus}(1974)}]{birpikus_eng}
\bibinfo{author}{\bibfnamefont{G.~L.} \bibnamefont{Bir}} \bibnamefont{and}
  \bibinfo{author}{\bibfnamefont{G.~E.} \bibnamefont{Pikus}},
  \emph{\bibinfo{title}{Symmetry and Deformational Effects in Semiconductors}}
  (\bibinfo{publisher}{Wiley, New York}, \bibinfo{year}{1974}).

\bibitem[{\citenamefont{Ivchenko et~al.}(1990)\citenamefont{Ivchenko,
  Lyanda-Geller, and Pikus}}]{ivchenko1990current}
\bibinfo{author}{\bibfnamefont{E.~L.} \bibnamefont{Ivchenko}},
  \bibinfo{author}{\bibfnamefont{Y.~B.} \bibnamefont{Lyanda-Geller}},
  \bibnamefont{and} \bibinfo{author}{\bibfnamefont{G.~E.} \bibnamefont{Pikus}},
  \bibinfo{journal}{Sov. Phys.-JETP} \textbf{\bibinfo{volume}{71}},
  \bibinfo{pages}{550} (\bibinfo{year}{1990}).

\bibitem[{\citenamefont{Khaetskii and Nazarov}(2001)}]{PhysRevB.64.125316}
\bibinfo{author}{\bibfnamefont{A.~V.} \bibnamefont{Khaetskii}}
  \bibnamefont{and} \bibinfo{author}{\bibfnamefont{Y.~V.}
  \bibnamefont{Nazarov}}, \bibinfo{journal}{Phys. Rev. B}
  \textbf{\bibinfo{volume}{64}}, \bibinfo{pages}{125316}
  (\bibinfo{year}{2001}).

\bibitem[{\citenamefont{Tsitsishvili et~al.}(2003)\citenamefont{Tsitsishvili,
  Baltz, and Kalt}}]{PhysRevB.67.205330}
\bibinfo{author}{\bibfnamefont{E.}~\bibnamefont{Tsitsishvili}},
  \bibinfo{author}{\bibfnamefont{R.~V.} \bibnamefont{Baltz}}, \bibnamefont{and}
  \bibinfo{author}{\bibfnamefont{H.}~\bibnamefont{Kalt}},
  \bibinfo{journal}{Phys. Rev. B} \textbf{\bibinfo{volume}{67}},
  \bibinfo{pages}{205330} (\bibinfo{year}{2003}).

\bibitem[{\citenamefont{Goupalov et~al.}(2003)\citenamefont{Goupalov,
  Lavallard, Lamouche, and Citrin}}]{Goupalov03_eng}
\bibinfo{author}{\bibfnamefont{S.~V.} \bibnamefont{Goupalov}},
  \bibinfo{author}{\bibfnamefont{P.}~\bibnamefont{Lavallard}},
  \bibinfo{author}{\bibfnamefont{G.}~\bibnamefont{Lamouche}}, \bibnamefont{and}
  \bibinfo{author}{\bibfnamefont{D.~S.} \bibnamefont{Citrin}},
  \bibinfo{journal}{Phys. Solid State} \textbf{\bibinfo{volume}{45}},
  \bibinfo{pages}{768} (\bibinfo{year}{2003}).

\bibitem[{\citenamefont{Glazov et~al.}(2014)\citenamefont{Glazov, Amand, Marie,
  Lagarde, Bouet, and Urbaszek}}]{Glazov2014}
\bibinfo{author}{\bibfnamefont{M.~M.} \bibnamefont{Glazov}},
  \bibinfo{author}{\bibfnamefont{T.}~\bibnamefont{Amand}},
  \bibinfo{author}{\bibfnamefont{X.}~\bibnamefont{Marie}},
  \bibinfo{author}{\bibfnamefont{D.}~\bibnamefont{Lagarde}},
  \bibinfo{author}{\bibfnamefont{L.}~\bibnamefont{Bouet}}, \bibnamefont{and}
  \bibinfo{author}{\bibfnamefont{B.}~\bibnamefont{Urbaszek}},
  \bibinfo{journal}{Phys. Rev. B} \textbf{\bibinfo{volume}{89}},
  \bibinfo{pages}{201302} (\bibinfo{year}{2014}).

\bibitem[{\citenamefont{Yu et~al.}(2018)\citenamefont{Yu, Liu, and
  Yao}}]{Yu_2018}
\bibinfo{author}{\bibfnamefont{H.}~\bibnamefont{Yu}},
  \bibinfo{author}{\bibfnamefont{G.-B.} \bibnamefont{Liu}}, \bibnamefont{and}
  \bibinfo{author}{\bibfnamefont{W.}~\bibnamefont{Yao}}, \bibinfo{journal}{2D
  Materials} \textbf{\bibinfo{volume}{5}}, \bibinfo{pages}{035021}
  (\bibinfo{year}{2018}).

\bibitem[{\citenamefont{Wang et~al.}(2015)\citenamefont{Wang, Bouet, Glazov,
  Amand, Ivchenko, Palleau, Marie, and Urbaszek}}]{Wang2015b}
\bibinfo{author}{\bibfnamefont{G.}~\bibnamefont{Wang}},
  \bibinfo{author}{\bibfnamefont{L.}~\bibnamefont{Bouet}},
  \bibinfo{author}{\bibfnamefont{M.~M.} \bibnamefont{Glazov}},
  \bibinfo{author}{\bibfnamefont{T.}~\bibnamefont{Amand}},
  \bibinfo{author}{\bibfnamefont{E.~L.} \bibnamefont{Ivchenko}},
  \bibinfo{author}{\bibfnamefont{E.}~\bibnamefont{Palleau}},
  \bibinfo{author}{\bibfnamefont{X.}~\bibnamefont{Marie}}, \bibnamefont{and}
  \bibinfo{author}{\bibfnamefont{B.}~\bibnamefont{Urbaszek}},
  \bibinfo{journal}{2D Mater.} \textbf{\bibinfo{volume}{2}},
  \bibinfo{pages}{034002} (\bibinfo{year}{2015}).

\bibitem[{\citenamefont{Durnev and Glazov}(2018)}]{DurnevUFN}
\bibinfo{author}{\bibfnamefont{M.~V.} \bibnamefont{Durnev}} \bibnamefont{and}
  \bibinfo{author}{\bibfnamefont{M.~M.} \bibnamefont{Glazov}},
  \bibinfo{journal}{Phys. Usp} \textbf{\bibinfo{volume}{61}},
  \bibinfo{pages}{825} (\bibinfo{year}{2018}).

\bibitem[{\citenamefont{Castellanos-Gomez
  et~al.}(2014)\citenamefont{Castellanos-Gomez, Buscema, Molenaar, Singh,
  Janssen, van~der Zant, and Steele}}]{Castellanos2014}
\bibinfo{author}{\bibfnamefont{A.}~\bibnamefont{Castellanos-Gomez}},
  \bibinfo{author}{\bibfnamefont{M.}~\bibnamefont{Buscema}},
  \bibinfo{author}{\bibfnamefont{R.}~\bibnamefont{Molenaar}},
  \bibinfo{author}{\bibfnamefont{V.}~\bibnamefont{Singh}},
  \bibinfo{author}{\bibfnamefont{L.}~\bibnamefont{Janssen}},
  \bibinfo{author}{\bibfnamefont{H.~S.~J.} \bibnamefont{van~der Zant}},
  \bibnamefont{and} \bibinfo{author}{\bibfnamefont{G.~A.}
  \bibnamefont{Steele}}, \bibinfo{journal}{2D Materials}
  \textbf{\bibinfo{volume}{1}}, \bibinfo{pages}{011002} (\bibinfo{year}{2014}).

\end{thebibliography}

\begin{thebibliography}{18}%
\makeatletter
\providecommand \@ifxundefined [1]{%
 \@ifx{#1\undefined}
}%
\providecommand \@ifnum [1]{%
 \ifnum #1\expandafter \@firstoftwo
 \else \expandafter \@secondoftwo
 \fi
}%
\providecommand \@ifx [1]{%
 \ifx #1\expandafter \@firstoftwo
 \else \expandafter \@secondoftwo
 \fi
}%
\providecommand \natexlab [1]{#1}%
\providecommand \enquote  [1]{``#1''}%
\providecommand \bibnamefont  [1]{#1}%
\providecommand \bibfnamefont [1]{#1}%
\providecommand \citenamefont [1]{#1}%
\providecommand \href@noop [0]{\@secondoftwo}%
\providecommand \href [0]{\begingroup \@sanitize@url \@href}%
\providecommand \@href[1]{\@@startlink{#1}\@@href}%
\providecommand \@@href[1]{\endgroup#1\@@endlink}%
\providecommand \@sanitize@url [0]{\catcode `\\12\catcode `\$12\catcode
  `\&12\catcode `\#12\catcode `\^12\catcode `\_12\catcode `\%12\relax}%
\providecommand \@@startlink[1]{}%
\providecommand \@@endlink[0]{}%
\providecommand \url  [0]{\begingroup\@sanitize@url \@url }%
\providecommand \@url [1]{\endgroup\@href {#1}{\urlprefix }}%
\providecommand \urlprefix  [0]{URL }%
\providecommand \Eprint [0]{\href }%
\providecommand \doibase [0]{http://dx.doi.org/}%
\providecommand \selectlanguage [0]{\@gobble}%
\providecommand \bibinfo  [0]{\@secondoftwo}%
\providecommand \bibfield  [0]{\@secondoftwo}%
\providecommand \translation [1]{[#1]}%
\providecommand \BibitemOpen [0]{}%
\providecommand \bibitemStop [0]{}%
\providecommand \bibitemNoStop [0]{.\EOS\space}%
\providecommand \EOS [0]{\spacefactor3000\relax}%
\providecommand \BibitemShut  [1]{\csname bibitem#1\endcsname}%
\let\auto@bib@innerbib\@empty
\bibitem [{\citenamefont {Carvalho}\ \emph {et~al.}(2017)\citenamefont
  {Carvalho}, \citenamefont {Wang}, \citenamefont {Mignuzzi}, \citenamefont
  {Roy}, \citenamefont {Terrones}, \citenamefont {Fantini}, \citenamefont
  {Crespi}, \citenamefont {Malard},\ and\ \citenamefont
  {Pimenta}}]{carvalho2017intervalley}%
  \BibitemOpen
  \bibfield  {author} {\bibinfo {author} {\bibfnamefont {B.~R.}\ \bibnamefont
  {Carvalho}}, \bibinfo {author} {\bibfnamefont {Y.}~\bibnamefont {Wang}},
  \bibinfo {author} {\bibfnamefont {S.}~\bibnamefont {Mignuzzi}}, \bibinfo
  {author} {\bibfnamefont {D.}~\bibnamefont {Roy}}, \bibinfo {author}
  {\bibfnamefont {M.}~\bibnamefont {Terrones}}, \bibinfo {author}
  {\bibfnamefont {C.}~\bibnamefont {Fantini}}, \bibinfo {author} {\bibfnamefont
  {V.~H.}\ \bibnamefont {Crespi}}, \bibinfo {author} {\bibfnamefont {L.~M.}\
  \bibnamefont {Malard}}, \ and\ \bibinfo {author} {\bibfnamefont {M.~A.}\
  \bibnamefont {Pimenta}},\ }\bibfield  {title} {\enquote {\bibinfo {title}
  {{Intervalley scattering by acoustic phonons in two-dimensional MoS$_2$
  revealed by double-resonance Raman spectroscopy}},}\ }\href@noop {}
  {\bibfield  {journal} {\bibinfo  {journal} {Nat. Commun.}\ }\textbf {\bibinfo
  {volume} {8}},\ \bibinfo {pages} {1} (\bibinfo {year} {2017})}\BibitemShut
  {NoStop}%
\bibitem [{\citenamefont {Kaasbjerg}\ \emph {et~al.}(2012)\citenamefont
  {Kaasbjerg}, \citenamefont {Thygesen},\ and\ \citenamefont
  {Jacobsen}}]{PhysRevB.85.115317}%
  \BibitemOpen
  \bibfield  {author} {\bibinfo {author} {\bibfnamefont {K.}~\bibnamefont
  {Kaasbjerg}}, \bibinfo {author} {\bibfnamefont {K.~S.}\ \bibnamefont
  {Thygesen}}, \ and\ \bibinfo {author} {\bibfnamefont {K.~W.}\ \bibnamefont
  {Jacobsen}},\ }\bibfield  {title} {\enquote {\bibinfo {title}
  {{Phonon-limited mobility in $n$-type single-layer MoS${}_{2}$ from first
  principles}},}\ }\href {\doibase 10.1103/PhysRevB.85.115317} {\bibfield
  {journal} {\bibinfo  {journal} {Phys. Rev. B}\ }\textbf {\bibinfo {volume}
  {85}},\ \bibinfo {pages} {115317} (\bibinfo {year} {2012})}\BibitemShut
  {NoStop}%
\bibitem [{\citenamefont {Liu}\ \emph {et~al.}(2013)\citenamefont {Liu},
  \citenamefont {Nestoklon}, \citenamefont {Cheng}, \citenamefont {Ivchenko},\
  and\ \citenamefont {Wu}}]{Liu2013}%
  \BibitemOpen
  \bibfield  {author} {\bibinfo {author} {\bibfnamefont {Z.}~\bibnamefont
  {Liu}}, \bibinfo {author} {\bibfnamefont {M.~O.}\ \bibnamefont {Nestoklon}},
  \bibinfo {author} {\bibfnamefont {J.~L.}\ \bibnamefont {Cheng}}, \bibinfo
  {author} {\bibfnamefont {E.~L.}\ \bibnamefont {Ivchenko}}, \ and\ \bibinfo
  {author} {\bibfnamefont {M.~W.}\ \bibnamefont {Wu}},\ }\bibfield  {title}
  {\enquote {\bibinfo {title} {Spin-dependent intravalley and intervalley
  electron-phonon scatterings in germanium},}\ }\href {\doibase
  10.1134/S1063783413080167} {\bibfield  {journal} {\bibinfo  {journal} {Phys.
  Solid State}\ }\textbf {\bibinfo {volume} {55}},\ \bibinfo {pages} {1619}
  (\bibinfo {year} {2013})}\BibitemShut {NoStop}%
\bibitem [{\citenamefont {Fallahazad}\ \emph {et~al.}(2016)\citenamefont
  {Fallahazad}, \citenamefont {Movva}, \citenamefont {Kim}, \citenamefont
  {Larentis}, \citenamefont {Taniguchi}, \citenamefont {Watanabe},
  \citenamefont {Banerjee},\ and\ \citenamefont
  {Tutuc}}]{PhysRevLett.116.086601}%
  \BibitemOpen
  \bibfield  {author} {\bibinfo {author} {\bibfnamefont {B.}~\bibnamefont
  {Fallahazad}}, \bibinfo {author} {\bibfnamefont {H.~C.~P.}\ \bibnamefont
  {Movva}}, \bibinfo {author} {\bibfnamefont {K.}~\bibnamefont {Kim}}, \bibinfo
  {author} {\bibfnamefont {S.}~\bibnamefont {Larentis}}, \bibinfo {author}
  {\bibfnamefont {T.}~\bibnamefont {Taniguchi}}, \bibinfo {author}
  {\bibfnamefont {K.}~\bibnamefont {Watanabe}}, \bibinfo {author}
  {\bibfnamefont {S.~K.}\ \bibnamefont {Banerjee}}, \ and\ \bibinfo {author}
  {\bibfnamefont {E.}~\bibnamefont {Tutuc}},\ }\bibfield  {title} {\enquote
  {\bibinfo {title} {{Shubnikov--de Haas Oscillations of High-Mobility Holes in
  Monolayer and Bilayer ${\mathrm{WSe}}_{2}$: Landau Level Degeneracy,
  Effective Mass, and Negative Compressibility}},}\ }\href {\doibase
  10.1103/PhysRevLett.116.086601} {\bibfield  {journal} {\bibinfo  {journal}
  {Phys. Rev. Lett.}\ }\textbf {\bibinfo {volume} {116}},\ \bibinfo {pages}
  {086601} (\bibinfo {year} {2016})}\BibitemShut {NoStop}%
\bibitem [{\citenamefont {Larentis}\ \emph {et~al.}(2018)\citenamefont
  {Larentis}, \citenamefont {Movva}, \citenamefont {Fallahazad}, \citenamefont
  {Kim}, \citenamefont {Behroozi}, \citenamefont {Taniguchi}, \citenamefont
  {Watanabe}, \citenamefont {Banerjee},\ and\ \citenamefont
  {Tutuc}}]{PhysRevB.97.201407}%
  \BibitemOpen
  \bibfield  {author} {\bibinfo {author} {\bibfnamefont {S.}~\bibnamefont
  {Larentis}}, \bibinfo {author} {\bibfnamefont {H.~C.~P.}\ \bibnamefont
  {Movva}}, \bibinfo {author} {\bibfnamefont {B.}~\bibnamefont {Fallahazad}},
  \bibinfo {author} {\bibfnamefont {K.}~\bibnamefont {Kim}}, \bibinfo {author}
  {\bibfnamefont {A.}~\bibnamefont {Behroozi}}, \bibinfo {author}
  {\bibfnamefont {T.}~\bibnamefont {Taniguchi}}, \bibinfo {author}
  {\bibfnamefont {K.}~\bibnamefont {Watanabe}}, \bibinfo {author}
  {\bibfnamefont {S.~K.}\ \bibnamefont {Banerjee}}, \ and\ \bibinfo {author}
  {\bibfnamefont {E.}~\bibnamefont {Tutuc}},\ }\bibfield  {title} {\enquote
  {\bibinfo {title} {{Large effective mass and interaction-enhanced Zeeman
  splitting of $K$-valley electrons in ${\mathrm{MoSe}}_{2}$}},}\ }\href
  {\doibase 10.1103/PhysRevB.97.201407} {\bibfield  {journal} {\bibinfo
  {journal} {Phys. Rev. B}\ }\textbf {\bibinfo {volume} {97}},\ \bibinfo
  {pages} {201407} (\bibinfo {year} {2018})}\BibitemShut {NoStop}%
\bibitem [{\citenamefont {Peng}\ \emph {et~al.}(2016)\citenamefont {Peng},
  \citenamefont {Zhang}, \citenamefont {Shao}, \citenamefont {Xu},
  \citenamefont {Zhang},\ and\ \citenamefont {Zhu}}]{C5RA19747C}%
  \BibitemOpen
  \bibfield  {author} {\bibinfo {author} {\bibfnamefont {B.}~\bibnamefont
  {Peng}}, \bibinfo {author} {\bibfnamefont {H.}~\bibnamefont {Zhang}},
  \bibinfo {author} {\bibfnamefont {H.}~\bibnamefont {Shao}}, \bibinfo {author}
  {\bibfnamefont {Y.}~\bibnamefont {Xu}}, \bibinfo {author} {\bibfnamefont
  {X.}~\bibnamefont {Zhang}}, \ and\ \bibinfo {author} {\bibfnamefont
  {H.}~\bibnamefont {Zhu}},\ }\bibfield  {title} {\enquote {\bibinfo {title}
  {{Thermal conductivity of monolayer MoS$_2${,} MoSe$_2${,} and WS$_2$:
  interplay of mass effect{,} interatomic bonding and anharmonicity}},}\ }\href
  {\doibase 10.1039/C5RA19747C} {\bibfield  {journal} {\bibinfo  {journal} {RSC
  Adv.}\ }\textbf {\bibinfo {volume} {6}},\ \bibinfo {pages} {5767} (\bibinfo
  {year} {2016})}\BibitemShut {NoStop}%
\bibitem [{\citenamefont {Robert}\ \emph {et~al.}(2020)\citenamefont {Robert},
  \citenamefont {Han}, \citenamefont {Kapuscinski}, \citenamefont {Delhomme},
  \citenamefont {Faugeras}, \citenamefont {Amand}, \citenamefont {Molas},
  \citenamefont {Bartos}, \citenamefont {Watanabe}, \citenamefont {Taniguchi},
  \citenamefont {Urbaszek}, \citenamefont {Potemski},\ and\ \citenamefont
  {Marie}}]{robert2020measurement}%
  \BibitemOpen
  \bibfield  {author} {\bibinfo {author} {\bibfnamefont {C.}~\bibnamefont
  {Robert}}, \bibinfo {author} {\bibfnamefont {B.}~\bibnamefont {Han}},
  \bibinfo {author} {\bibfnamefont {P.}~\bibnamefont {Kapuscinski}}, \bibinfo
  {author} {\bibfnamefont {A.}~\bibnamefont {Delhomme}}, \bibinfo {author}
  {\bibfnamefont {C.}~\bibnamefont {Faugeras}}, \bibinfo {author}
  {\bibfnamefont {T.}~\bibnamefont {Amand}}, \bibinfo {author} {\bibfnamefont
  {M.~R.}\ \bibnamefont {Molas}}, \bibinfo {author} {\bibfnamefont
  {M.}~\bibnamefont {Bartos}}, \bibinfo {author} {\bibfnamefont
  {K.}~\bibnamefont {Watanabe}}, \bibinfo {author} {\bibfnamefont
  {T.}~\bibnamefont {Taniguchi}}, \bibinfo {author} {\bibfnamefont
  {B.}~\bibnamefont {Urbaszek}}, \bibinfo {author} {\bibfnamefont
  {M.}~\bibnamefont {Potemski}}, \ and\ \bibinfo {author} {\bibfnamefont
  {X.}~\bibnamefont {Marie}},\ }\bibfield  {title} {\enquote {\bibinfo {title}
  {{Measurement of the spin-forbidden dark excitons in MoS$_2$ and MoSe$_2$
  monolayers}},}\ }\href@noop {} {\bibfield  {journal} {\bibinfo  {journal}
  {Nat. Commun.}\ }\textbf {\bibinfo {volume} {11}},\ \bibinfo {pages} {4037}
  (\bibinfo {year} {2020})}\BibitemShut {NoStop}%
\bibitem [{\citenamefont {Yu}\ \emph {et~al.}(2018)\citenamefont {Yu},
  \citenamefont {Liu},\ and\ \citenamefont {Yao}}]{Yu_2018}%
  \BibitemOpen
  \bibfield  {author} {\bibinfo {author} {\bibfnamefont {H.}~\bibnamefont
  {Yu}}, \bibinfo {author} {\bibfnamefont {G.-B.}\ \bibnamefont {Liu}}, \ and\
  \bibinfo {author} {\bibfnamefont {W.}~\bibnamefont {Yao}},\ }\bibfield
  {title} {\enquote {\bibinfo {title} {{Brightened spin-triplet interlayer
  excitons and optical selection rules in van der Waals heterobilayers}},}\
  }\href {\doibase 10.1088/2053-1583/aac065} {\bibfield  {journal} {\bibinfo
  {journal} {2D Mater.}\ }\textbf {\bibinfo {volume} {5}},\ \bibinfo {pages}
  {035021} (\bibinfo {year} {2018})}\BibitemShut {NoStop}%
\bibitem [{\citenamefont {Wo\ifmmode~\acute{z}\else \'{z}\fi{}niak}\ \emph
  {et~al.}(2020)\citenamefont {Wo\ifmmode~\acute{z}\else \'{z}\fi{}niak},
  \citenamefont {Faria~J.}, \citenamefont {Seifert}, \citenamefont {Chaves},\
  and\ \citenamefont {Kunstmann}}]{wozniak2020exciton}%
  \BibitemOpen
  \bibfield  {author} {\bibinfo {author} {\bibfnamefont {T.}~\bibnamefont
  {Wo\ifmmode~\acute{z}\else \'{z}\fi{}niak}}, \bibinfo {author} {\bibfnamefont
  {Paulo~E.}\ \bibnamefont {Faria~J.}}, \bibinfo {author} {\bibfnamefont
  {G.}~\bibnamefont {Seifert}}, \bibinfo {author} {\bibfnamefont
  {A.}~\bibnamefont {Chaves}}, \ and\ \bibinfo {author} {\bibfnamefont
  {J.}~\bibnamefont {Kunstmann}},\ }\bibfield  {title} {\enquote {\bibinfo
  {title} {{Exciton $g$ factors of van der Waals heterostructures from
  first-principles calculations}},}\ }\href {\doibase
  10.1103/PhysRevB.101.235408} {\bibfield  {journal} {\bibinfo  {journal}
  {Phys. Rev. B}\ }\textbf {\bibinfo {volume} {101}},\ \bibinfo {pages}
  {235408} (\bibinfo {year} {2020})}\BibitemShut {NoStop}%
\bibitem [{\citenamefont {Yugova}\ \emph {et~al.}(2009)\citenamefont {Yugova},
  \citenamefont {Glazov}, \citenamefont {Ivchenko},\ and\ \citenamefont
  {Efros}}]{yugova09}%
  \BibitemOpen
  \bibfield  {author} {\bibinfo {author} {\bibfnamefont {I.~A.}\ \bibnamefont
  {Yugova}}, \bibinfo {author} {\bibfnamefont {M.~M.}\ \bibnamefont {Glazov}},
  \bibinfo {author} {\bibfnamefont {E.~L.}\ \bibnamefont {Ivchenko}}, \ and\
  \bibinfo {author} {\bibfnamefont {Al.~L.}\ \bibnamefont {Efros}},\ }\bibfield
   {title} {\enquote {\bibinfo {title} {{Pump-probe Faraday rotation and
  ellipticity in an ensemble of singly charged quantum dots}},}\ }\href
  {\doibase 10.1103/PhysRevB.80.104436} {\bibfield  {journal} {\bibinfo
  {journal} {Phys. Rev. B}\ }\textbf {\bibinfo {volume} {80}},\ \bibinfo {eid}
  {104436} (\bibinfo {year} {2009})}\BibitemShut {NoStop}%
\bibitem [{\citenamefont {Holler}\ \emph {et~al.}(2022)\citenamefont {Holler},
  \citenamefont {Selig}, \citenamefont {Kempf}, \citenamefont {Zipfel},
  \citenamefont {Nagler}, \citenamefont {Katzer}, \citenamefont {Katsch},
  \citenamefont {Ballottin}, \citenamefont {Mitioglu}, \citenamefont
  {Chernikov}, \citenamefont {Christianen}, \citenamefont {Sch\"uller},
  \citenamefont {Knorr},\ and\ \citenamefont {Korn}}]{PhysRevB.105.085303}%
  \BibitemOpen
  \bibfield  {author} {\bibinfo {author} {\bibfnamefont {J.}~\bibnamefont
  {Holler}}, \bibinfo {author} {\bibfnamefont {M.}~\bibnamefont {Selig}},
  \bibinfo {author} {\bibfnamefont {M.}~\bibnamefont {Kempf}}, \bibinfo
  {author} {\bibfnamefont {J.}~\bibnamefont {Zipfel}}, \bibinfo {author}
  {\bibfnamefont {P.}~\bibnamefont {Nagler}}, \bibinfo {author} {\bibfnamefont
  {M.}~\bibnamefont {Katzer}}, \bibinfo {author} {\bibfnamefont
  {F.}~\bibnamefont {Katsch}}, \bibinfo {author} {\bibfnamefont {M.~V.}\
  \bibnamefont {Ballottin}}, \bibinfo {author} {\bibfnamefont {A.~A.}\
  \bibnamefont {Mitioglu}}, \bibinfo {author} {\bibfnamefont {A.}~\bibnamefont
  {Chernikov}}, \bibinfo {author} {\bibfnamefont {P.~C.~M.}\ \bibnamefont
  {Christianen}}, \bibinfo {author} {\bibfnamefont {C.}~\bibnamefont
  {Sch\"uller}}, \bibinfo {author} {\bibfnamefont {A.}~\bibnamefont {Knorr}}, \
  and\ \bibinfo {author} {\bibfnamefont {T.}~\bibnamefont {Korn}},\ }\bibfield
  {title} {\enquote {\bibinfo {title} {Interlayer exciton valley polarization
  dynamics in large magnetic fields},}\ }\href {\doibase
  10.1103/PhysRevB.105.085303} {\bibfield  {journal} {\bibinfo  {journal}
  {Phys. Rev. B}\ }\textbf {\bibinfo {volume} {105}},\ \bibinfo {pages}
  {085303} (\bibinfo {year} {2022})}\BibitemShut {NoStop}%
\bibitem [{\citenamefont {Korm\'anyos}\ \emph {et~al.}(2014)\citenamefont
  {Korm\'anyos}, \citenamefont {Z\'olyomi}, \citenamefont {Drummond},\ and\
  \citenamefont {Burkard}}]{Kormanyos2014}%
  \BibitemOpen
  \bibfield  {author} {\bibinfo {author} {\bibfnamefont {A.}~\bibnamefont
  {Korm\'anyos}}, \bibinfo {author} {\bibfnamefont {V.}~\bibnamefont
  {Z\'olyomi}}, \bibinfo {author} {\bibfnamefont {N.~D.}\ \bibnamefont
  {Drummond}}, \ and\ \bibinfo {author} {\bibfnamefont {G.}~\bibnamefont
  {Burkard}},\ }\bibfield  {title} {\enquote {\bibinfo {title} {{Spin-Orbit
  Coupling, Quantum Dots, and Qubits in Monolayer Transition Metal
  Dichalcogenides}},}\ }\href {\doibase 10.1103/PhysRevX.4.011034} {\bibfield
  {journal} {\bibinfo  {journal} {Phys. Rev. X}\ }\textbf {\bibinfo {volume}
  {4}},\ \bibinfo {pages} {011034} (\bibinfo {year} {2014})}\BibitemShut
  {NoStop}%
\bibitem [{\citenamefont {Khaetskii}\ and\ \citenamefont
  {Nazarov}(2001)}]{PhysRevB.64.125316}%
  \BibitemOpen
  \bibfield  {author} {\bibinfo {author} {\bibfnamefont {A.~V.}\ \bibnamefont
  {Khaetskii}}\ and\ \bibinfo {author} {\bibfnamefont {Y.~V.}\ \bibnamefont
  {Nazarov}},\ }\bibfield  {title} {\enquote {\bibinfo {title} {Spin-flip
  transitions between {Z}eeman sublevels in semiconductor quantum dots},}\
  }\href {\doibase 10.1103/PhysRevB.64.125316} {\bibfield  {journal} {\bibinfo
  {journal} {Phys. Rev. B}\ }\textbf {\bibinfo {volume} {64}},\ \bibinfo
  {pages} {125316} (\bibinfo {year} {2001})}\BibitemShut {NoStop}%
\bibitem [{\citenamefont {Tong}\ \emph {et~al.}(2020)\citenamefont {Tong},
  \citenamefont {Chen}, \citenamefont {Xiao}, \citenamefont {Yu},\ and\
  \citenamefont {Yao}}]{Tong_2020}%
  \BibitemOpen
  \bibfield  {author} {\bibinfo {author} {\bibfnamefont {Q.}~\bibnamefont
  {Tong}}, \bibinfo {author} {\bibfnamefont {M.}~\bibnamefont {Chen}}, \bibinfo
  {author} {\bibfnamefont {F.}~\bibnamefont {Xiao}}, \bibinfo {author}
  {\bibfnamefont {H.}~\bibnamefont {Yu}}, \ and\ \bibinfo {author}
  {\bibfnamefont {W.}~\bibnamefont {Yao}},\ }\bibfield  {title} {\enquote
  {\bibinfo {title} {Interferences of electrostatic moir{\'{e}} potentials and
  bichromatic superlattices of electrons and excitons in transition metal
  dichalcogenides},}\ }\href {\doibase 10.1088/2053-1583/abd006} {\bibfield
  {journal} {\bibinfo  {journal} {2D Mater.}\ }\textbf {\bibinfo {volume}
  {8}},\ \bibinfo {pages} {025007} (\bibinfo {year} {2020})}\BibitemShut
  {NoStop}%
\bibitem [{\citenamefont {Bir}\ and\ \citenamefont
  {Pikus}(1974)}]{birpikus_eng}%
  \BibitemOpen
  \bibfield  {author} {\bibinfo {author} {\bibfnamefont {G.~L.}\ \bibnamefont
  {Bir}}\ and\ \bibinfo {author} {\bibfnamefont {G.~E.}\ \bibnamefont
  {Pikus}},\ }\href@noop {} {\emph {\bibinfo {title} {Symmetry and
  Deformational Effects in Semiconductors}}}\ (\bibinfo  {publisher} {Wiley,
  New York},\ \bibinfo {year} {1974})\BibitemShut {NoStop}%
\bibitem [{\citenamefont {Ivchenko}\ \emph {et~al.}(1990)\citenamefont
  {Ivchenko}, \citenamefont {Lyanda-Geller},\ and\ \citenamefont
  {Pikus}}]{ivchenko1990current}%
  \BibitemOpen
  \bibfield  {author} {\bibinfo {author} {\bibfnamefont {E.~L.}\ \bibnamefont
  {Ivchenko}}, \bibinfo {author} {\bibfnamefont {Yu.~B.}\ \bibnamefont
  {Lyanda-Geller}}, \ and\ \bibinfo {author} {\bibfnamefont {G.~E.}\
  \bibnamefont {Pikus}},\ }\bibfield  {title} {\enquote {\bibinfo {title}
  {Current of thermalized spin-oriented photocarriers},}\ }\href@noop {}
  {\bibfield  {journal} {\bibinfo  {journal} {Sov. Phys.-JETP}\ }\textbf
  {\bibinfo {volume} {71}},\ \bibinfo {pages} {550} (\bibinfo {year}
  {1990})}\BibitemShut {NoStop}%
\bibitem [{\citenamefont {Song}\ and\ \citenamefont {Dery}(2013)}]{Song2013}%
  \BibitemOpen
  \bibfield  {author} {\bibinfo {author} {\bibfnamefont {Y.}~\bibnamefont
  {Song}}\ and\ \bibinfo {author} {\bibfnamefont {H.}~\bibnamefont {Dery}},\
  }\bibfield  {title} {\enquote {\bibinfo {title} {{Transport Theory of
  Monolayer Transition-Metal Dichalcogenides through Symmetry}},}\ }\href
  {\doibase 10.1103/PhysRevLett.111.026601} {\bibfield  {journal} {\bibinfo
  {journal} {Phys. Rev. Lett.}\ }\textbf {\bibinfo {volume} {111}},\ \bibinfo
  {pages} {026601} (\bibinfo {year} {2013})}\BibitemShut {NoStop}%
\bibitem [{\citenamefont {Pearce}\ \emph {et~al.}(2016)\citenamefont {Pearce},
  \citenamefont {Mariani},\ and\ \citenamefont {Burkard}}]{PhysRevB.94.155416}%
  \BibitemOpen
  \bibfield  {author} {\bibinfo {author} {\bibfnamefont {A.~J.}\ \bibnamefont
  {Pearce}}, \bibinfo {author} {\bibfnamefont {E.}~\bibnamefont {Mariani}}, \
  and\ \bibinfo {author} {\bibfnamefont {G.}~\bibnamefont {Burkard}},\
  }\bibfield  {title} {\enquote {\bibinfo {title} {Tight-binding approach to
  strain and curvature in monolayer transition-metal dichalcogenides},}\ }\href
  {\doibase 10.1103/PhysRevB.94.155416} {\bibfield  {journal} {\bibinfo
  {journal} {Phys. Rev. B}\ }\textbf {\bibinfo {volume} {94}},\ \bibinfo
  {pages} {155416} (\bibinfo {year} {2016})}\BibitemShut {NoStop}%
\end{thebibliography}

\let\addcontentsline\oldaddcontentsline
\makeatletter
\renewcommand\tableofcontents{%
    \@starttoc{toc}%
}
\makeatother
\renewcommand{\i}{{\rm i}}


\onecolumngrid
\vspace{\columnsep}
\begin{center}
\newpage
\makeatletter
{\large\bf{Supplemental Material to\\``\@title''}}
\makeatother
\end{center}
\vspace{\columnsep}

The Supplementary Material includes the following topics:

\hypersetup{linktoc=page}
\tableofcontents
\vspace{\columnsep}

\counterwithin{figure}{section}
\renewcommand{\thepage}{S\arabic{page}}
\renewcommand{\theequation}{S\arabic{equation}}
\renewcommand{\thesection}{S\arabic{section}}
\renewcommand{\thefigure}{S\arabic{figure}}

\setcounter{page}{1}
\setcounter{section}{0}
\setcounter{equation}{0}
\setcounter{figure}{0}

\section{Details of calculation of hole intervalley scattering time}

In this section we present details of the calculation of the hole intervalley scattering rate in the interlayer exciton based on Eq. (3) in the main text. We note that the scattering can be equally considered as a two step process consisting of phonon emission and exciton scattering be exchange interaction or as the phonon induced scattering between excitonic states mixed by the long-range exchange interaction.


The electron intervalley scattering in the considered mechanism takes place within the MoSe$_2$ monolayer (ML). So we consider a single isolated ML to derive a Hamiltonian of the electron-phonon interaction. The common point group of the wave vectors $K_+$ and $K_-$ is $C_{3h}$. We chose the center of transformations at the hollow center of a hexagon. The electron states in $K_\pm$ valleys of the lower (upper) subband of the conduction band transform according to $K_{11}$ ($K_{9}$) and $K_{12}$ ($K_{10}$) irreducible representations, respectively.

We consider the electron spin conserving intervalley scattering from the lower to the upper subband. To derive the selection rules for the electron-phonon interaction, we consider the scattering from $K_+$ to $K_-$ valley, as shown in Fig.~2(c) in the main text. The selection rules for the opposite case follow from  time reversal symmetry. The scattering requires emission of a $K_-$ phonon or absorption of a $K_+$ phonon; the selection rules for these processes are the same. TA phonons (polarized in the ML plane) at $K_\pm$ valleys transform according to $K_3$ and $K_2$ irreducible representation, respectively~\cite{carvalho2017intervalley}.

Multiplication of the representations of final ($K_{10}^*=K_9$), initial ($K_{11}$) electron representation and phonon representation ($K_2^*=K_3$) reads
\begin{equation}
  K_9\otimes K_3\otimes K_{11} = K_2.
\end{equation}
This demonstrates that this scattering is forbidden exactly between the $K$ points. However, it is allowed between the states at the wave vectors $\bm K_++\bm q_+$ and $\bm K_-+\bm q_-$ in the first order in $\bm q_+$ and $\bm q_-$~\cite{PhysRevB.85.115317}. Since the time reversal relates $K_+$ and $K_-$ valleys, and the matrix elements of the spin independent scattering of the direct and reverse processes are the same, the matrix element is proportional to the components of $\bm q=\bm q_+-\bm q_-$~\cite{Liu2013}.

The components $q_x\pm\i q_y$ transform according to $K_2$ and $K_3$ irreducible representations, respectively. As a result, the Hamiltonian of the electron-phonon interaction has the form
\begin{equation}
  \mathcal H_{\text{e-ph}}=\sum_{\bm q_\pm,s_z}\sqrt{\frac{\hbar}{2\rho\Omega_qA}}\Xi q_- a_{\bm K_-+\bm q_-,s_z}^\dag a_{\bm K_++\bm q_+,s_z}\left(b_{\bm K_+-\bm q}+b_{\bm K_-+\bm q}^\dag\right)+\text{H.c.},
\end{equation}
where $a_{\bm k,s_z}$ ($a_{\bm k,s_z}^\dag$) are the electron annihilation (creation) operators for the state with the wave vector $\bm k$ and spin $s_z$ in the conduction band. Taking into account that
\begin{equation}
  \blue{\hat{\tau}_\pm}=\sum_{\bm q,s_z}a_{\bm K_\pm+\bm q,s_z}^\dag a_{\bm K_\mp+\bm q,s_z}
\end{equation}
this expression is equivalent to Eq.~(4) in the main text.

The selection rules and the form of the Hamiltonian can be understood from the consideration of the angular momenta of electrons and phonons. Neglecting spin, the electron has an orbital angular momentum $\pm1$ at $K_\pm$ valleys, respectively. The chiral TA phonons at $K_\pm$ points have angular momenta $\mp 1$, respectively. Taking into account $C_{3h}$ symmetry of the intervalley scattering, the angular momentum should be conserved modulo 3. As a result, the electron scattering from $K_\pm$ to $K_\mp$ valley requires the factor of $q_\mp$.

To calculate the matrix elements of the electron-phonon interaction we consider exciton wave functions of the form
\begin{equation}
  \label{eq:Psi}
  \Psi_{\alpha}(\bm\rho,\bm R)=\frac{\sqrt{2}}{\pi a_B l}\exp\left(-\frac{\rho}{a_B}-\frac{R^2}{2l^2}\right)\xi_{\alpha},
\end{equation}
where $\alpha=i,f,a$ denotes initial, final, and auxiliary states, respectively, $\bm R=(\bm r_e+\bm r_h)/2$ is the coordinate of the exciton center of mass with the hole coordinate $\bm r_h$ and the electron, $m_e$, and hole, $m_h$, masses assumed to be equal for simplicity, $\xi_{\alpha}$ is the spinor describing the exciton valley and spin state. This form of the wave functions corresponds to exciton localization in a parabolic potential at a length much larger than the exciton Bohr radius, $l\gg a_B$. \blue{We have checked that consideration of additional localized states increases the scattering rate by no more than 40\%.}

With these wave functions for low temperatures, $k_BT\ll g_h\mu_BB_{\text{res}}$, we obtain the matrix element of electron scattering with emission of a phonon with the wave vector $\bm q$:
\begin{equation}
  \label{eq:H_ph}
  \left|\left\langle a_{\bm q}\middle|\mathcal H_{e-ph}^{\bm q}\middle|i\right\rangle\right|=\sqrt{\frac{\hbar}{2\rho\Omega_{q}A}}\Xi q\exp\left[-\left(\frac{ql}{2}\right)^2\right].
\end{equation}


The calculation of the matrix element of the long-range electron hole exchange interaction after Eq.~(5) in the main text for the wave functions from Eq.~\eqref{eq:Psi} gives
\begin{equation}
  \label{eq:H_ex}
  \left\langle f\middle|\mathcal H_{exch}\middle|a\right\rangle=\frac{\sqrt{\pi}\hbar\sqrt{\Gamma_0^a\Gamma_0^f}}{2 kl}.
\end{equation}


Finally to calculate the hole intervalley scattering rate we consider the parabolic dispersion of TA phonons in the vicinity of $K$ points and obtain the phonon density of states
\begin{equation}
  \label{eq:D0}
  D_0(E)=\frac{A M}{2\pi\hbar^2}\theta(E-\hbar\Omega_0),
\end{equation}
with $\theta(\varepsilon)$ being the Heaviside step function. Combining this with the matrix elements of the electron-phonon and exchange interactions, Eqs.~\eqref{eq:H_ph} and~\eqref{eq:H_ex}, from Eq.~(3) in the main text we obtain Eq.~(6) in the main text. It describes a narrow resonance in $\tau_{\text{res}}$ at the field $B\approx B_{\text{res}}$ with the maximum value given by Eq.~(7) in the main text.

To estimate the scattering rate we use the following material parameters: exciton mass $m=1.25m_0$~\cite{PhysRevLett.116.086601,PhysRevB.97.201407}, phonon effective mass determined from the fit of the phonon dispersion $M=0.1m_p$ with $m_p$ being the proton mass~\cite{C5RA19747C}, localization energy $E_{\text{loc}}=10$~meV, intervalley deformation potential $\Xi=5.9$~eV~\cite{PhysRevB.85.115317}, areal density of MoSe$_2$ ML $\rho=4.5\cdot10^{-7}$~g/cm$^2$, intervalley exciton energy $\hbar\omega_0=1.4$~eV, background dielectric constant $\varkappa_b=6$, spin orbit splitting of MoSe$_2$ ML conduction band without contribution of the short range exchange interaction $\Delta_c=20$~meV~\cite{robert2020measurement}, relative exciton oscillator strengths $\Gamma_0^a\Gamma_0^f/\Gamma_0^2=2.7\cdot10^{-5}$ for H-type HS and $0.9\cdot10^{-5}$ for R-type HS~\cite{Yu_2018} with $\hbar\Gamma_0=1$~meV being the homogeneous exciton linewidth in MoSe$_2$ ML, and electron valley g factor $g_e^v=\mp3.6$ for H-type and R-type HS, respectively~\cite{wozniak2020exciton}. With these parameters we obtain reasonable scattering times $\tau_{\text{res}}=12$~ns and $20$~ns for H-type and R-type HS, respectively, as given in the main text.

Another estimation can be made for the exciton radiative lifetime~\cite{yugova09}
\begin{equation}
  \frac{1}{\tau_R}=\frac{4k^3e^2|p_{cv}^2|}{3\varkappa_b\hbar\omega_0^2m_0^2}\left|\int\Psi(0,{\bm R})\d{\bm R}\right|^2=\frac{8}{3}\Gamma_0^f(kl)^2.
\end{equation}
Using the same parameters we obtain $\tau_R=84$~ns and $\tau_R=37$~ns for H-type and R-type HS, respectively, which agrees with the timescale of the PL decay~\cite{PhysRevB.105.085303}.

\section{Discussion of alternative mechanisms}

In this section we make estimations for two other possible mechanisms of the hole intervalley scattering. They demonstrate that the suggested scattering mechanism involving spin-conserving electron intervalley scattering and long-range electron hole exchange interaction is the dominant one.

\subsection{Admixture mechanism}

The admixture mechanism is based on the hole spin-orbit interaction in addition to the spin-conserving hole-phonon interaction. The corresponding scattering rate can be calculated as
\begin{equation}
  \label{eq:admixture}
  \frac{1}{\tau_{\text{ad}}}=\frac{2\pi}{\hbar}\sum_{a,\bm q}\left|\frac{\left\langle f_{\bm q}'\middle|\mathcal H_{SO}\middle|a_{\bm q}'\right\rangle\left\langle a_{\bm q}'\middle|\mathcal H_{h-ph}\middle|i\right\rangle}{E_f-E_a}\right|^2\delta(E_i-E_f-\hbar\Omega_{\bm q}),
\end{equation}
which is similar to Eq.~(3) in the main text, but with spin-orbit interaction instead of the exchange interaction and different phonon and auxiliary states.

The spin-orbit interaction requires breaking of the horizontal mirror reflection symmetry of the ML, which naturally happens for heterobilayers. In the given valley its Hamiltonian can be written as~\cite{Kormanyos2014}
\begin{equation}
  \mathcal H_{SO}=\lambda_{\alpha\beta}k_\alpha s_\beta,
\end{equation}
where $\lambda_{\alpha\beta}$ are the spin-orbit coupling constants and $k_\alpha$ and $s_\beta$ are the hole momentum and spin components. The matrix elements can be estimated as $\left\langle f_{\bm q}'\middle|\mathcal H_{SO}\middle|a_{\bm q}'\right\rangle\sim\lambda/a_B$, where $\lambda$ is the typical spin-orbit coupling constant.

The hole valley-magnetophonon resonance at 24.2~T corresponds to the energy of the chiral TA phonon of MoSe$_2$ and the ZA phonon of WSe$_2$. Similarly to the electron, the hole spin-conserving intervalley scattering is symmetry forbidden exactly between the corners of the Brillouin zone. Therefore the matrix element of the hole-phonon interaction in analogy with Eq.~\eqref{eq:H_ph} can be estimated as $\left\langle a_{\bm q}'\middle|\mathcal H_{h-ph}\middle|i\right\rangle\sim\sqrt{\hbar^2/(\rho g_h\mu_B B_{\text{res}}A)}\Xi_{\text{ad}}/a_B$ with hole spin-conserving intervalley deformation potential $\Xi_{\text{ad}}$.

Finally, the van-Vleck cancellation in the admixture mechanism requires a modification of the exciton wave function by the magnetic field~\cite{PhysRevB.64.125316}. This brings an additional factor $(a_B/l_B)^2$ to the combined matrix element with $l_B=\sqrt{\hbar/(eB)}$ being the magnetic length.

Now from Eq.~\eqref{eq:admixture} using the phonon density of states, Eq.~\eqref{eq:D0}, the scattering rate in the admixture mechanism can be estimated as
\begin{equation}
  \frac{1}{\tau_{\text{ad}}}\sim\frac{M\Xi_{\text{ad}}^2\lambda^2e^2B_{\text{res}}}{\hbar^3\rho\Delta_v^2g_h\mu_B},
\end{equation}
where we the energy difference $|E_f-E_a|$ is replaced with the valence band spin orbit splitting $\Delta_v$. For an estimation we use the spin-orbit coupling constant $\lambda=0.5\cdot10^{-3}$~eV$\cdot$\AA~\cite{Kormanyos2014}, which agrees with the order of magnitude of the built-in electric field in HS~\cite{Tong_2020}. We also note that the interaction of a hole with a TA phonon of MoSe$_2$ is suppressed by the hole localization in the WSe$_2$ ML, while spin-conserving interaction with a ZA phonon requires  breaking of the mirror reflection symmetry. Therefore, we take a smaller intervalley deformation potential $\Xi_{\text{ad}}=1$~eV. For all the other parameters as above and the valence band spin-orbit splitting $\Delta_v=300$~meV we obtain the hole intervalley scattering time for the admixture mechanism $\tau_{\text{ad}}\sim6$~ms. This is approximately three orders of magnitude larger, than for the mechanism described in the main text.

\subsection{Direct spin-phonon coupling}

Direct hole spin-flip scattering exactly between the $K$ valleys is allowed for a ZA phonon of WSe$_2$ for the given point symmetry. However it is forbidden by time reversal symmetry~\cite{birpikus_eng,ivchenko1990current,Song2013}. As a result, the intervalley hole-phonon interaction involves the second power of $q\sim1/a_B$. In addition, the scattering between two localized states related by time reversal symmetry involves van Vleck cancellation. Similarly to the admixture mechanism this gives an additional factor of $(a_B/l_B)^2$ to the scattering matrix element, which can be estimated as
\begin{equation}
  \left\langle f_{\bm q}''\middle|\mathcal H_{h-ph}\middle|i\right\rangle\sim\sqrt{\frac{\hbar^2}{\rho g_h\mu_B B_{\text{res}}A}}\frac{\Xi_{\text{dir}}}{a_B^2}\left(\frac{a_B}{l_B}\right)^2
\end{equation}
with the spin-flip intervalley deformation potential $\Xi_{\text{dir}}$.

Using the phonon density of states, Eq.~\eqref{eq:D0}, we obtain an estimation for the scattering rate
\begin{equation}
  \frac{1}{\tau_{\text{dir}}}\sim\frac{M\Xi_{\text{dir}}^2e^2B_{\text{res}}}{\hbar^3\rho g_h\mu_B}
\end{equation}
For an estimation we take $\Xi_{\text{dir}}=19$~meV$\cdot$\AA~\cite{PhysRevB.94.155416} and obtain $\tau_{\text{dir}}\sim50~\mu$s. This is approximately four orders of magnitude larger than for the mechanism described in the main text.


\section{Model of PL polarization}

To describe the PL DOP as a function of magnetic field we consider only the lowest subband of the conduction band and the upper subband of the valence band. We take into account exciton generation, radiative and non radiative recombinations, as well as electron and hole valley relaxations, as shown in Fig.~2(a,b) in the main text.

For R-type HS the following set of kinetic equations describes the exciton dynamics:
\begin{subequations}
  \begin{equation}
    \frac{\d n_{\Uparrow\uparrow}}{\d t}=G-\frac{n_{\Uparrow\uparrow}}{\tau_R}-\frac{n_{\Uparrow\uparrow}}{\tau_h(B)}\exp\left(-\frac{g_h\mu_B B}{k_B T}\right)+\frac{n_{\Downarrow\uparrow}}{\tau_h(B)}-\frac{n_{\Uparrow\uparrow}}{\tau_e(B)}+\frac{n_{\Uparrow\downarrow}}{\tau_e(B)}\exp\left(-\frac{g_e\mu_B B}{k_B T}\right),
  \end{equation}
  \begin{equation}
    \frac{\d n_{\Downarrow\downarrow}}{\d t}=G-\frac{n_{\Downarrow\downarrow}}{\tau_R}-\frac{n_{\Downarrow\downarrow}}{\tau_h(B)}+\frac{n_{\Uparrow\downarrow}}{\tau_h(B)}\exp\left(-\frac{g_h\mu_B B}{k_B T}\right)-\frac{n_{\Downarrow\downarrow}}{\tau_e(B)}\exp\left(-\frac{g_e\mu_B B}{k_B T}\right)+\frac{n_{\Downarrow\uparrow}}{\tau_e(B)},
  \end{equation}
  \begin{equation}
    \frac{\d n_{\Uparrow\downarrow}}{\d t}=G-\frac{n_{\Uparrow\downarrow}}{\tau_{NR}}-\frac{n_{\Uparrow\downarrow}}{\tau_h(B)}\exp\left(-\frac{g_h\mu_B B}{k_B T}\right)+\frac{n_{\Downarrow\downarrow}}{\tau_h(B)}-\frac{n_{\Uparrow\downarrow}}{\tau_e(B)}\exp\left(-\frac{g_e\mu_B B}{k_B T}\right)+\frac{n_{\Uparrow\uparrow}}{\tau_e(B)},
  \end{equation}
  \begin{equation}
    \frac{\d n_{\Downarrow\uparrow}}{\d t}=G-\frac{n_{\Downarrow\uparrow}}{\tau_{NR}}-\frac{n_{\Downarrow\uparrow}}{\tau_h(B)}+\frac{n_{\Uparrow\uparrow}}{\tau_h(B)}\exp\left(-\frac{g_h\mu_B B}{k_B T}\right)-\frac{n_{\Downarrow\uparrow}}{\tau_e(B)}+\frac{n_{\Downarrow\downarrow}}{\tau_e(B)}\exp\left(-\frac{g_e\mu_B B}{k_B T}\right).
  \end{equation}
\end{subequations}
Here $\Uparrow$, $\Downarrow$ and $\uparrow$, $\downarrow$ denote the hole (empty space in the valence band, strictly speaking) and electron valley, respectively; $n$ with the corresponding subscript denotes occupancy of the excitonic state; the generation rate $G$ is assumed to be the same for all excitonic states because of the nonresonant exciton pumping; $\tau_R$ and $\tau_{NR}$ are the radiative and nonradiative recombination times, respectively. The electron and hole valley relaxation times $\tau_h(B)$ and $\tau_e(B)$ are introduced in the main text. For the H-type HS $\tau_e(B)$ and $\tau_e(B)\exp\left[-g_e\mu_BB/(k_BT)\right]$ should be exchanged.

From the solution of these equations in the steady state, PL DOP can be calculated as
\begin{equation}
  P_{PL}=\pm\frac{n_{\Downarrow\downarrow}-n_{\Uparrow\uparrow}}{n_{\Downarrow\downarrow}+n_{\Uparrow\uparrow}},
\end{equation}
for R-type and H-type HS, respectively. For the fits shown in Fig.~1(e,f) for R(H)-type HS the following parameters are used: $\tau_{\text{res}}/\tau_R=4.7(0.87)$, $\Delta B=0.6$~T, $\tau_{NR}/\tau_R=36(1.4)$, $\tau_h^{(0)}/\tau_R=2(6)$, $\tau_e^{(0)}/\tau_R=5.7(0.4)$. One can see that these fits not only reproduce the resonant enhancement of DOP at the field $B_{\text{res}}$, but also nicely describe DOP in the whole range of magnetic fields.

\end{document}